\pdfoutput=1

\documentclass[11pt,twoside,a4paper,cmspaper,final,collab]{cms-tdr}
\def\svnVersion{365277}\def\cmsCernNoTag{CERN-EP/2016-004}\def\cmsCernDate{\today}\def\cmsMessage{Published in the Journal of High Energy Physics as \href{http://dx.doi.org/10.1007/JHEP07(2016)027}{\doi{10.1007/JHEP07(2016)027.}}}

\begin{document}\cmsNoteHeader{SUS-14-015}

\hyphenation{had-ron-i-za-tion}
\hyphenation{cal-or-i-me-ter}
\hyphenation{de-vices}
\RCS$HeadURL: svn+ssh://svn.cern.ch/reps/tdr2/papers/SUS-14-015/trunk/SUS-14-015.tex $
\RCS$Id: SUS-14-015.tex 344794 2016-05-30 13:23:28Z bargassa $

\newlength\cmsFigWidth
\ifthenelse{\boolean{cms@external}}{\setlength\cmsFigWidth{0.85\columnwidth}}{\setlength\cmsFigWidth{0.4\textwidth}}
\ifthenelse{\boolean{cms@external}}{\providecommand{\cmsLeft}{top}}{\providecommand{\cmsLeft}{left}}
\ifthenelse{\boolean{cms@external}}{\providecommand{\cmsRight}{bottom}}{\providecommand{\cmsRight}{right}}

\newcommand{\tauh}{\ensuremath{\tau_\mathrm{h}}\xspace}
\newcommand{\mt}{\ensuremath{M_{\mathrm{T}}}\xspace}
\newcommand{\met}{\ensuremath{E^\mathrm{miss}_\mathrm{T}}\xspace}
\newcommand{\ptl}{\ensuremath{p_{\mathrm T}(\ell)}\xspace}
\newcommand{\mw}{\ensuremath{M_{{\mathrm T}2}^{\PW}}\xspace}
\newcommand{\ptj}{\ensuremath{p_{\mathrm T}(\mathrm{j}_1)}\xspace}
\newcommand{\ptb}{\ensuremath{p_{\mathrm T}(\cPqb_1)}\xspace}
\newcommand{\HTfrac}{\ensuremath{H^\text{frac}_{\mathrm T}}\xspace}
\newcommand{\drLB}{\ensuremath{\Delta R (\ell, \cPqb_1)}\xspace}
\newcommand{\dfMJ}{\ensuremath{\Delta \phi(\mathrm{j}_{1,2}, \ptvecmiss)}\xspace}
\newcommand{\ChiHad}{\ensuremath{\chi^2_{\text{had}}}\xspace}
\newcommand{\njets}{\ensuremath{N(\text{jets})}\xspace}
\newcommand{\bjet}{\ensuremath{\text{b jet}}\xspace}
\newcommand{\bjets}{\ensuremath{\text{b jets}}\xspace}
\newcommand{\mtb}{\ensuremath{M(\text{3\,jet})}\xspace}
\newcommand{\mlb}{\ensuremath{M(\ell\cPqb)}\xspace}
\newcommand{\MTlb}{\ensuremath{M^\prime_{\ell \cPqb}}\xspace}

\newcommand{\MTTwo}{\ensuremath{M_{\mathrm{T2}}}\xspace}
\newcommand{\MTTwoS}[2]{\ensuremath{M_{\mathrm{T2}}^{#1 #2}}\xspace}
\newcommand{\vecpt}{\ensuremath{\vec{p}_{\mathrm T}}\xspace}
\newcommand{\SpecvecptMiss}[1]{\ensuremath{\vec{p}_{\mathrm{T#1}}^{\,\text{miss}}}\xspace}
\newcommand{\mz}{\ensuremath{M_{\Z}}\xspace}
\newcommand{\mll}{\ensuremath{M_{\ell^+\ell^-}}\xspace}

\newcommand{\ttl}{\ensuremath{\ttbar\to1\ell}}
\newcommand{\ttll}{\ensuremath{\ttbar\to\ell\ell}}
\newcommand{\ttdl}{\ensuremath{\ttbar\to\ell\ell}}
\newcommand{\wjets}{\ensuremath{\PW+}\text{jets}\xspace}
\newcommand{\SFpre}{\ensuremath{SF_{\ell\ell}}\xspace}
\newcommand{\SFpost}{\ensuremath{SF_{0}}\xspace}
\newcommand{\lsp}{\ensuremath{\PSGczDo}\xspace}
\newcommand{\chg}{\ensuremath{\PSGcpmDo}\xspace}
\renewcommand{\sTop}{\PSQt}
\newcommand{\stp}{\ensuremath{\kern0.2em\sTop_{1}}\xspace}
\newcommand{\dm}{\ensuremath{\Delta m}\xspace}
\newcommand{\mmplane}{\ensuremath{ (  {{m}}(\stp),   {{m}}(\lsp) )}\xspace}
\newcommand\pfMET{\textsc{pfMET}\xspace}
\newcommand\pfJET{\textsc{pfJET}\xspace}
\newcommand{\ch}{\checkmark}
\newcommand{\df}{\ensuremath{\Delta \phi}}
\newcommand{\dr}{\ensuremath{\Delta R}}
\newcommand{\bs}{0.038\linewidth}
\newcommand{\bsb}{0.048\linewidth}
\newcommand{\bsx}{0.060\linewidth}
\newcommand{\bss}{0.028\linewidth}

 \cmsNoteHeader{SUS-14-015}
 \title{Search for direct pair production of scalar top quarks in the single- and dilepton channels in proton-proton collisions at $\sqrt{s}=8\TeV$}

\date{\today}

\abstract{
Results are reported from a search for the top squark \stp, the lighter
of the two supersymmetric partners of the top quark.
The data sample corresponds to 19.7\fbinv of proton-proton collisions at $\sqrt{s} = 8\TeV$
collected with the CMS detector at the LHC.
The search targets $\stp\to \cPqb
\PSGcpmDo$ and $\stp\to \cPqt^{(*)} \PSGczDo$
decay modes, where $\PSGcpmDo$ and $\PSGczDo$ are the
lightest chargino and neutralino, respectively. The reconstructed final
state consists of jets, b jets, missing transverse energy, and either one or two
leptons. Leading backgrounds are determined from data. No significant
excess in data is observed above the expectation from standard model
processes. The results exclude a region of the two-dimensional plane of
possible \stp and \lsp masses. The highest excluded \stp and \lsp masses
are about 700\GeV and 250\GeV, respectively.
}

\hypersetup{%
pdfauthor={CMS Collaboration},%
pdftitle={Search for direct pair production of scalar top quarks in the single- and dilepton channels in proton-proton collisions at sqrt(s) = 8 TeV},%
pdfsubject={CMS},%
pdfkeywords={CMS, physics, SUSY}}

\maketitle

 \section{Introduction}
\label{s:Stop}

Theories of supersymmetry (SUSY) predict the existence of a scalar
partner for each standard model (SM) left-handed and right-handed
fermion. When the symmetry is broken, the scalar partners acquire a mass
different from their SM counterparts, the mass splitting between scalar
mass eigenstates being dependent on the mass of the SM fermion. Because
of the large mass of the top quark, the splitting between its chiral
supersymmetric partners is potentially the largest among all
supersymmetric quarks (squarks). As a result the lighter supersymmetric
scalar partner of the top quark, the top squark (\stp), could be the
lightest squark. The search for a low mass top squark is of particular
interest following the discovery of a Higgs
boson~\cite{HgDisc1,HgDisc2,HgDisc3}, as a top squark with a mass in the
TeV range would contribute substantially to the cancellation of the
divergent loop corrections to the Higgs boson mass. SUSY scenarios with
a neutralino (\lsp) as lightest supersymmetric particle (LSP) and a
nearly degenerate-mass \stp provide one theoretically possible way to
produce the observed relic abundance of dark matter~\cite{djoua,carena};
this further motivates the search for the \stp at the LHC.

\begin{figure}[hbt]
\centering
\includegraphics[width=\linewidth]{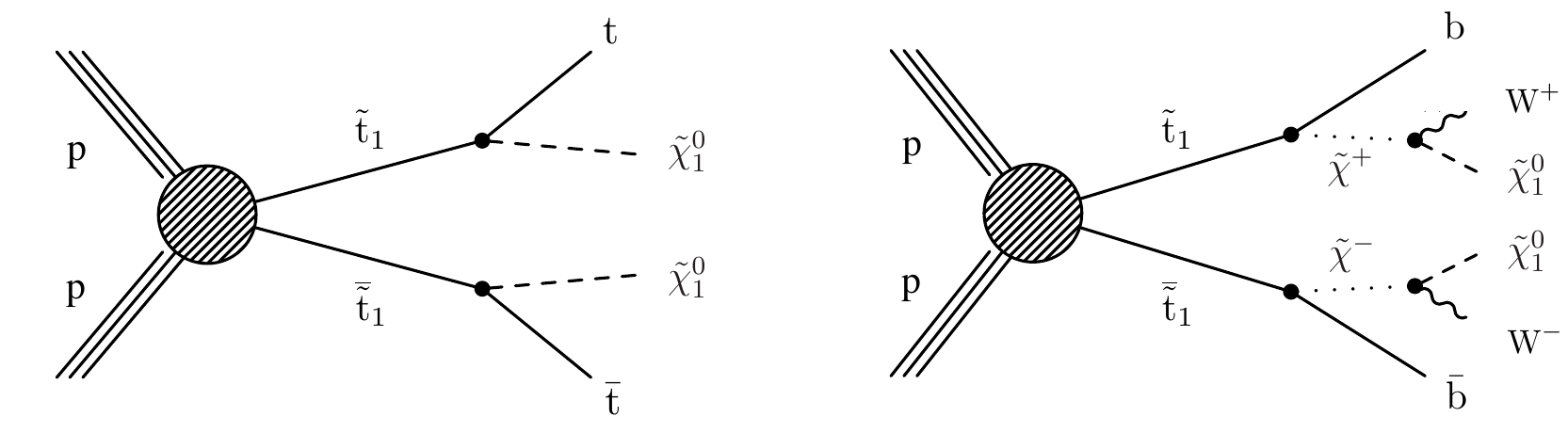}
\caption{
\label{fig:diagram}
Top squark direct pair production at the LHC. Left: tt decay mode. Right: bbWW decay mode.}
\end{figure}

In this paper we report two searches for direct top squark pair
production with the CMS detector at $\sqrt{s}=8$\TeV with integrated
luminosities of 19.5\fbinv and 19.7\fbinv. Each search is based on the
two decay modes shown in Fig.~\ref{fig:diagram}. The decay modes and the
nomenclature we will use to refer to them are as follows:
\begin{equation*}\begin{aligned}
\Pp\Pp&\to\sTop_{1}\PASQt_{1}\to \cPqt^{(*)}\cPaqt^{(*)}\PSGczDo\PSGczDo &&\text{(the ``tt'' decay mode);}\\
\Pp\Pp&\to\sTop_{1}\PASQt_{1}\to
\cPqb\cPaqb\PSGcpDo\PSGcmDo \to \cPqb\cPaqb \PW^{+(*)}\PW^{-(*)} \PSGczDo\PSGczDo &&\text{(the ``bbWW'' decay mode).}
\end{aligned}\end{equation*}
The tt and bbWW events both contain bottom quark jets (henceforth called
\bjets) and may contain charged leptons and neutrinos from $\PW^{(*)}$
decay. The search strategies are therefore tailored to require either
one lepton or two leptons, as well as at least one \bjet and a minimum
amount of transverse momentum imbalance.
Throughout this paper the term ``lepton'' refers only to $\Pe^\pm$ and
$\mu^\pm$. Previous searches for low mass top squarks in leptonic final
states have been conducted by the D0, CDF, CMS, and ATLAS
collaborations~\cite{Abazov:2003wt,Abazov:2012cz,Abazov:2007ak,Aaltonen:2010uf,Acosta:2003ys,
Aad:2015pfx,MtPub}.

As shown in Table~\ref{tab:StopDecay}, we categorize the decays of the
\stp as 2-body or 3-body processes and as a function of the masses of
the involved particles. In all cases we take the lightest neutralino
\lsp to be the LSP. For each decay mode we fix the corresponding \stp
branching fraction to unity; the search is in all other respects
designed to be as independent as possible of the details of any specific
SUSY model. We explore a range of signal mass points for each decay mode
under consideration. In the decay mode tt, the unknown masses are those
of the \stp and the \lsp, while in the case of bbWW, a third unknown is
the mass of the lightest chargino ($\PSGcpmDo$). In the
latter case we consider three possible mass assignments, labeled by the
parameter $x=0.25$, 0.50, 0.75; $x$ is defined by
\begin{equation}
\label{e:defx}
m(\PSGcpmDo) =
m(\PSGczDo) + x  \big[ m(\kern0.2em\sTop_1) - m(\PSGczDo) \big].
\end{equation}

\begin{table}[!ht]
\topcaption{
Kinematic conditions for the \stp decay modes explored in this paper.}
\centering
		\renewcommand{\arraystretch}{1.5}
        \begin{tabular}{lll}
            \hline
 Kinematic conditions &       Type of decay &  Decay mode \\
            \hline

 $m(\cPqb) + m(\PW) + m(\PSGczDo) \leq m(\kern0.2em\sTop_1)$ 	& \multirow{2}{*}{3-body decays (tt)}  & \multirow{2}{*}{$\sTop_1 (\to \cPqt^* \PSGczDo) \to \cPqb \PW \PSGczDo$}  \\
and $m(\kern0.2em\sTop_1) < m(\cPqt) + m(\PSGczDo)$		&				 &				\\
            \hline
$m(\cPqt) + m(\PSGczDo) \leq m(\kern0.2em\sTop_1)$		& 2-body decays (tt) &  $\sTop_1 \to \cPqt \PSGczDo$, with $\cPqt \to \cPqb \PW$ \\
           \hline
$m(\cPqb) + m(\PW) + m(\PSGczDo) \leq m(\kern0.2em\sTop_1)$ 	& \multirow{2}{*}{2-body decays (bbWW)} & \multirow{2}{*}{$\sTop_1 \to \cPqb \PSGcpmDo$, with $\PSGcpmDo \to \PW^{(*)} \PSGczDo$} \\
and  $m(\PSGczDo) < m(\PSGcpmDo) < m(\kern0.2em\sTop_1) - m(\cPqb)$  & 			& 				\\
            \hline
        \end{tabular}
\label{tab:StopDecay}
\end{table}

In this paper we expand the result of our previous search in the
single-lepton final state~\cite{MtPub} by improving key aspects of the
signal selection. Since the SM background dominates the signal by
several orders of magnitude and often has similar distributions for
individual discriminating variables, a multivariate approach has been
developed to exploit differences in the correlations among
discriminating variables for signal and SM background. The background
determination method has also been improved compared to
Ref.~\cite{MtPub} in order to better control and correct the tail of the
key transverse mass distribution. In addition to the single lepton
search, we also report on a search in the dilepton mode, where the key
discriminating variable is an \MTTwo\ variable ~\cite{Burns:2008va}. The
final result is based on a combination of the single lepton and dilepton
searches.

\section{The CMS detector}
\label{s:det}

The central feature of the CMS apparatus is a superconducting solenoid
that provides an axial magnetic field of 3.8\unit{T} for charged-particle
tracking. Trajectories of charged particles are measured by a silicon
pixel and strip tracker, covering $0 < \phi < 2\pi$ in azimuth and
$\abs{\eta}<2.5$, where the pseudorapidity $\eta$ is defined as $\eta =
-\ln[\tan (\theta/2)]$; $\theta$ is the polar angle of the trajectory of
the particle with respect to the counterclockwise beam direction. A
crystal electromagnetic calorimeter and a brass/scintillator hadronic
calorimeter surround the tracking detectors. The calorimeter measures
the energy and direction of electrons, photons, and hadronic jets. Muons
are measured in gas-ionization detectors embedded in the steel
flux-return yoke outside the solenoid. The detector is nearly hermetic,
allowing for momentum balance measurements in the plane transverse to
the beam axis. Events are selected online by a two-level trigger
system~\cite{hlt}. A more detailed description of the CMS detector can
be found in Ref.~\cite{CMSdet}.

\section{Samples, triggers, and reconstruction algorithms}
\label{s:str}

\subsection{Samples and trigger requirements}
\label{s:Trigger}

Events used for this search are selected initially by single-lepton and
dilepton triggers. For the single-electron final state, the online
selection requires the electron be isolated and have transverse momentum
$\pt>27$\GeV; in subsequent offline analysis the reconstructed electron
\pt is required to exceed 30\GeV. For the single-muon final state, two
triggers are used, which both require $\abs{\eta(\mu)}< 2.1$: a purely
leptonic trigger requiring an isolated muon with $\pt>24$\GeV; and an
additional mixed trigger requiring an isolated muon of $\pt>17$\GeV
together with three jets, each having $\pt>30$\GeV. The first trigger
suffices for muons whose offline reconstructed \pt exceeds 26\GeV,
while the second trigger allows the analysis to use muons with
reconstructed \pt as low as 20\GeV; the additional jets are required in
the analysis in any case. The dilepton triggers require either $\Pe\Pe$,
$\mu\mu$, or $\Pe\mu$ pairs. In each case, one lepton must satisfy $\pt
> 17$\GeV and the other lepton must satisfy $\pt > 8$\GeV. Trigger
efficiencies are measured in data and applied to simulated events. The
integrated luminosity, after data quality requirements, is $19.5 \pm
0.5\fbinv$ for the single-lepton states and $19.7 \pm 0.5\fbinv$ for
the dilepton final states~\cite{CMS:2013gfa}.

The SM background processes of relevance to this analysis are \ttbar,
\wjets, $\cPZ/\gamma^* \to \ell^+ \ell^-$ (denoted Drell--Yan, or DY),
single top, diboson, triboson, and \ttbar$+$ boson(s). They are simulated
by the \MADGRAPH~\cite{MG} (v5.1.3.30) and \POWHEG~\cite{powheg} event
generators, with CTEQ6L1~\cite{cteq6l1} and CT10~\cite{ct10} parton density
functions (PDF) respectively.
Simulated event samples with signal mass points chosen on a grid of \mmplane
values are generated, where the mass of the \stp varies between 100 and
1000\GeV, and the mass of the \lsp varies between 0 and 700\GeV; as mentioned
in Section \ref{s:Stop} (see Eq.~\eqref{e:defx}), three different mass
hierarchies are considered for the bbWW decay mode. The generation of signal
samples is performed using \MADGRAPH with CTEQ6L1 PDF.
Parton shower and hadronization are simulated using \PYTHIA~\cite{pythia}
(v6.4.26 for background and v6.4.22 for signal) with the tune Z2$^*$ \cite{Z2}.
All simulated events are propagated through the CMS detector model either
with the \GEANTfour package~\cite{geant}, or, in the case of the signal
samples, with a fast parametric simulation \cite{fastsim}.
The next-to-leading-order (NLO) plus the next-to-leading-log (NLL) cross
sections for top squark pair production are calculated with
\PROSPINO~\cite{prospino,xs1,xs2,xs3,xs4,xs5}.

To ensure agreement with data, simulated events are weighted so that the
distribution of the number of proton-proton interactions per beam
crossing agrees with that seen in data; they are additionally weighted
by the trigger efficiency and the lepton identification and isolation
efficiencies. For simulated \ttbar samples, a \pt-dependent weight is
applied to match the shape of the
$\rd\sigma(\Pp\Pp\to\ttbar+\mathrm{X})/\rd\pt$ distribution observed in
data. Signal events are weighted to account for the effect of initial
state radiation~\cite{MtPub}.

\subsection{Object reconstruction}
\label{s:Reco}

In this search, all particle candidates are reconstructed with the
particle-flow (PF) algorithm~\cite{PfJetMet1,PfJetMet2}, and additional
criteria are then applied to select electrons, muons, jets, and \bjets;
the criteria are applied to both collision data and simulation samples.

The identification and measurement of the \pt of muons uses information
provided by the silicon detector and the muon system~\cite{MuonID}. We
require the muon to have a `tight' identification~\cite{MuonID} with
pseudorapidity $\abs{\eta} < 2.1$ and $\abs{\eta} < 2.4$ for the single-lepton
and dilepton searches, respectively. The identification and energy
measurement of the electrons uses information provided by the tracker
and the electromagnetic calorimeter. Electron candidates are
reconstructed in the tracker with the Gaussian-sum filter
algorithm~\cite{EleID}. We require the electron to have a `medium'
identification~\cite{EleID} with pseudorapidity $\abs{\eta}< 1.44$ and
$\abs{\eta}< 2.5$ for the single-lepton and dilepton searches, respectively.
Both muon and electron identification demand that the lepton be isolated
from the hadronic components of the event. We define an isolation
variable for the leptons based on a scalar sum of transverse momenta,
$\mathcal{P} \equiv \sum \abs{\ptvec}$, where the sum is taken over
all PF candidates within a cone about the lepton of $\Delta R \equiv
\sqrt{\smash[b]{(\Delta \phi)^2+(\Delta \eta)^2}} = 0.3$, excluding the transverse
momentum of the lepton itself, $\pt(\ell)$. In the single-lepton search
we impose an upper limit on the absolute isolation, $\mathcal{P}< 5$\GeV;
for both searches we impose an upper limit on the relative isolation
$\mathcal{P}/\pt(\ell) < 0.15$.

Jets are constructed by clustering all the PF candidates with the
anti-\kt jet clustering algorithm~\cite{Cacciari:2008gp}, using a
distance parameter $R=0.5$.
Contamination from additional \Pp\Pp~interactions (pileup) is mitigated by
discarding charged PF candidates that are incompatible with having
originated from the estimated proton-proton collision point. The average
pileup energy associated with neutral hadrons is computed event-by-event
and subtracted from the jet energy and from the energy used when
computing lepton isolation, i.e., a measure of the activity around the
lepton. The energy subtracted is the average pileup energy per unit area
(in $\Delta \eta \times \Delta \Phi$) times the jet or isolation cone
area~\cite{pu1,pu2}.
Candidate jets must be separated from selected leptons by $\Delta R
>0.4$. Relative and absolute jet energy corrections are applied to the
raw jet momenta to establish a uniform jet response in $\abs{\eta}$ and
a calibrated response in jet \pt. We require the jets pass $\pt> 30$\GeV
and $\abs{\eta} < 2.4$. To tag jets originating from the hadronization
of b quarks, we utilize the combined secondary vertex algorithm at its
`medium' operating point~\cite{btag} with a corresponding efficiency for
b jets of 65\% and a mistag rate for light jets of 1\%. Scale factors
are applied to simulation samples to reproduce the efficiencies measured
in the data.

As the decays of \stp are expected to yield neutralinos and neutrinos in
their decay chain, genuine missing transverse momentum is expected in
the final state of signal events. We define the missing transverse
momentum by a sum over the transverse momenta of all PF candidates,
$\ptvecmiss \equiv -\sum\ptvec$. All calibration
corrections~\cite{MET8tev} have been applied to candidates used in the
sum. The magnitude of the \ptvecmiss vector is denoted by $\MET\equiv
\abs{\ptvecmiss}$. We reject events where known detector effects or noise lead
to anomalously large \met values.

\section{Single-lepton search}
\label{s:1lep}

In the single-lepton search, we consider only final states containing
one lepton (e or $\mu$ only) and several jets.

\subsection{Event Selection}
\label{s:1lepSel}

The preselection criteria are defined as follows:
\begin{itemize}
\item Exactly one identified and isolated lepton satisfying $\pt(\mu)> 20$\GeV or $\pt(\Pe)> 30$\GeV;
\item A veto is applied against the presence of a second lepton by requiring
that no additional isolated tracks or hadronically decaying $\tau$ lepton (\tauh)
candidates~\cite{MtPub} are present;
\item The number of jets and number of \bjets must satisfy $\njets\geq 4$ and $N(\bjets)\geq 1$;
\item $\MET>80\GeV$;
\item $\mt>100\GeV$.
\end{itemize}
The transverse mass variable is defined by $\mt \equiv \sqrt{\smash[b]{2 \met
\pt(\ell) (1-\cos \Delta\phi )}}$, where $\pt(\ell)$ is the
transverse momentum of the selected lepton and $\Delta\phi$ is the
angular difference between the lepton $\ptvec(\ell)$ and \ptvecmiss.  The
requirement on this variable suppresses events in which the source of
the lepton and \ptvecmiss is $\PW^\pm$ decay.

\begin{figure*}[htb]
\includegraphics[width=0.5\textwidth]{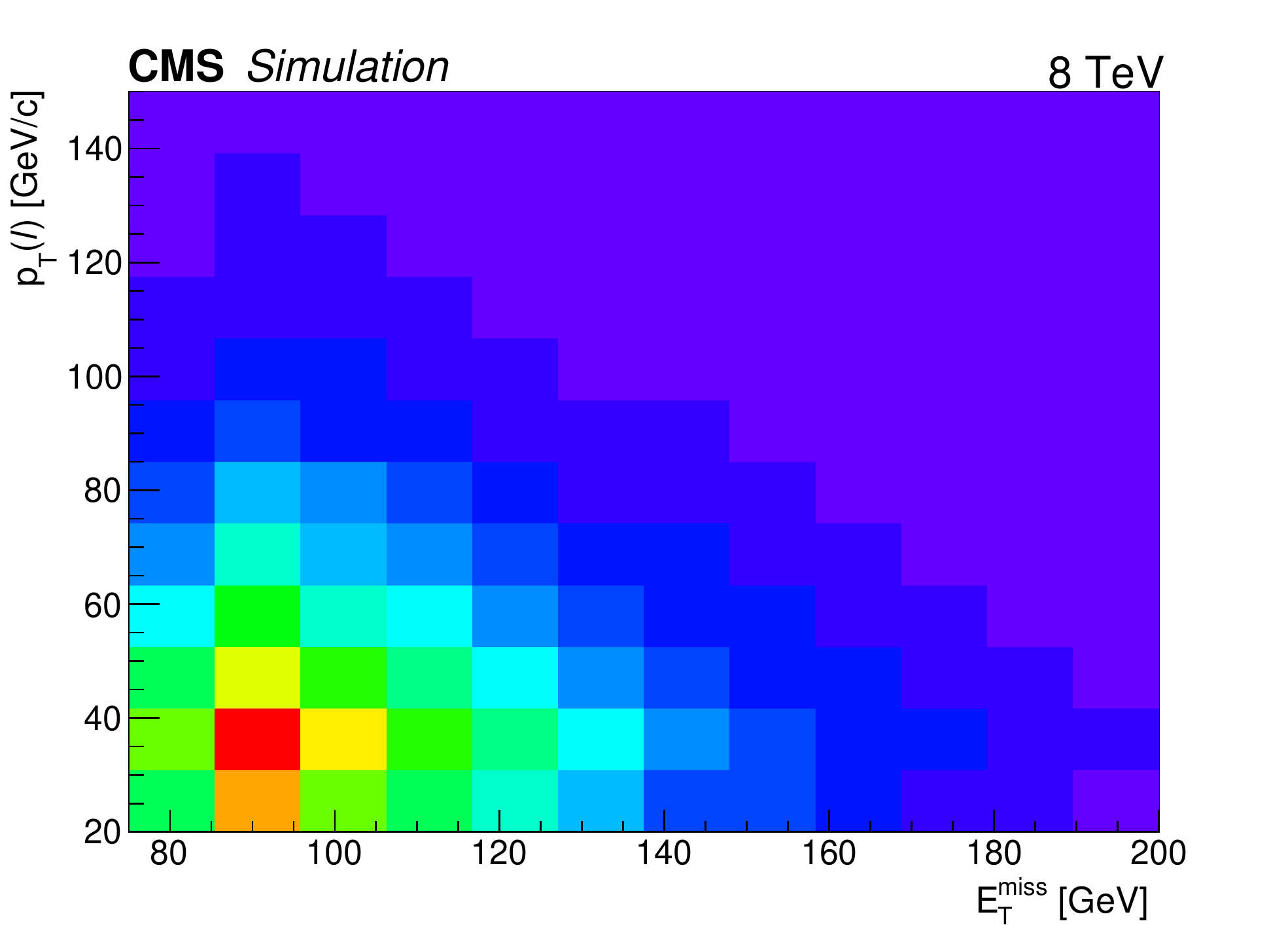}
\includegraphics[width=0.5\textwidth]{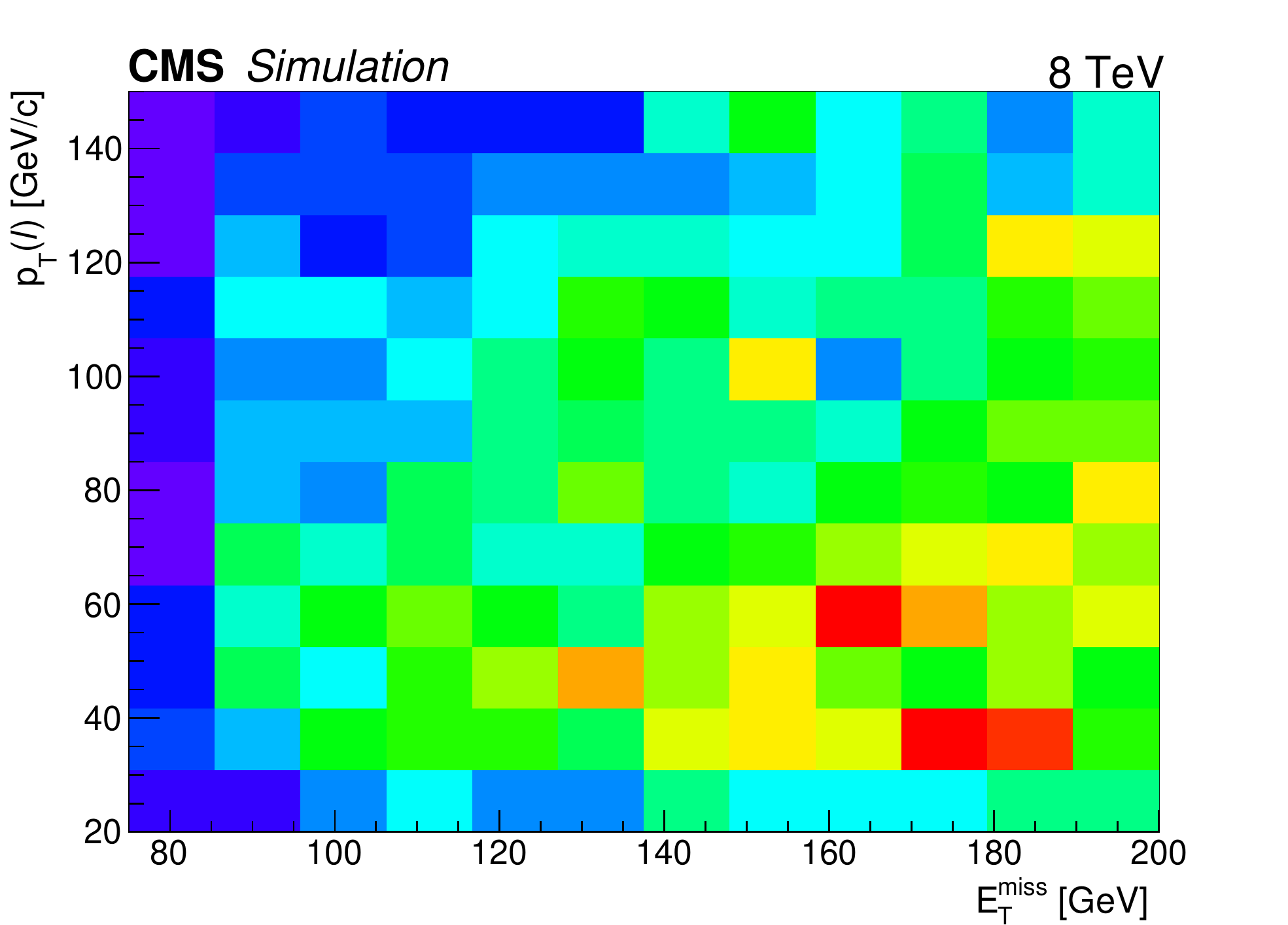}
\caption{Distribution of the transverse momentum of the lepton versus the
missing transverse energy at the preselection, for the simulated \ttbar background (left) and
for the bbWW decay mode ($x=0.50$) of the signal with m(\stp)-m(\lsp) $\geq$
625 GeV (right).}
\label{fig:CorrVar}
\end{figure*}

At the preselection level, the \ttbar and \wjets backgrounds represent
90\% and 7\%, respectively, of the total expected background (see
Section~\ref{s:1lepBckg}). For the signal selection, we use a boosted
decision tree (BDT)~\cite{RefBDT} to take advantage of the correlations
among variables that discriminate between signal and background;
Fig.~\ref{fig:CorrVar} illustrates how a pair of kinematic variables
correlate differently for a background process and signal.
Compared to the approach of Ref.~\cite{MtPub}, the signal selection is
characterized mainly by the use of new variables, and a systematic
search for the most reduced set of best-performing variables to be used
as input to the BDT. Furthermore, because the discriminating power of
each input varies across the \mmplane mass plane, the latter is
partitioned and a unique BDT is trained in each partition. The full list
of variables (not all used in every BDT) is given below:

\begin{itemize}

\item \met\!: The presence of missing transverse momentum signals the
possible production of a stable unseen object, such as the \lsp.

\item \ptl: The correlation between the missing transverse momentum
\met\ and the lepton transverse momentum \ptl differs between signal,
where genuine \met is due to two missing objects (\lsp), and \ttbar and
\wjets backgrounds where the \met is due to a single missing object
($\nu$).

\item \njets\!, \ptj\!, \ptb\!: These describe the multiplicity of selected jets and
 the \pt of the highest \pt jet and highest \pt \bjet, respectively.

\item \mw: The distribution of this variable shows an edge at the top
quark mass for \ttbar events where both W bosons decay leptonically and
one of the leptons is lost.

It is defined by minimizing the following over possible momentum vectors
$\vec{p}_{\mathrm{T}1}$ and $\vec{p}_{\mathrm{T}2}$:
\begin{equation}
\label{eq:MT2W_def}
 \mw = \min \left\{ \text{$M_x$ consistent with:} \left[ \begin{aligned}
\vec{p}_{{\mathrm T}1} + \vec{p}_{{\mathrm T}2} =  \ptvecmiss,~\left(p_1 + p_\ell \right)^2 \equiv p_2^2 = m(\PW)^2, \\
 p_1^2 = 0, ~\left(p_1 + p_\ell +  {p_{\cPqb}}_{1}  \right)^2 = \left(p_2 + {p_{\cPqb}}_{2} \right)^2 =M_x^2
\end{aligned}
\right] \right\}.
\end{equation}
Here $p_1$ is the momentum of the neutrino associated with a
successfully reconstructed lepton in one $\PW\to\ell\nu$ decay, and $p_2$ corresponds to an
unreconstructed \PW\ whose two decay products (the lost lepton and the
neutrino) escape detection. The momenta ${p_{\cPqb}}_{1}$ and ${p_{\cPqb}}_{2}$ are of
the  \bjets with the highest (leading) and second-highest (sub-leading) \pt values, respectively.
Including \mw in the BDT reduces the contribution of
the \ttbar dilepton background.

\item \HT:  The scalar sum $\HT\equiv\sum\abs{\ptvec}$, summed over all jets with
$\pt>30\GeV$, characterizes the hadronic component of the event. A related
variable \HTfrac is defined by $\HTfrac\equiv \sum^\prime\abs{\ptvec}/\HT$,
where the terms in the numerator are restricted to jets of
$\pt>30\GeV$ that lie in the same hemisphere as \ptvecmiss.

\item \drLB, \dfMJ: Two topological variables, \drLB and \dfMJ, are defined as follows:
$\Delta R$ is the distance between the lepton and the leading \bjet; and
$\Delta \phi$ is the minimal angular difference
between the \ptvecmiss vector and either
the leading or sub-leading jet.

\item $\ChiHad$:  To characterize the kinematics of \ttbar events
we build a $\chi^2$ variable comparing the invariant masses of the
three- and two-jet systems to the mass of the top quark and W boson,
respectively. It is defined as:
\begin{equation}
 \ChiHad = \frac{(M_{\mathrm{j}1\mathrm{j}2\mathrm{j}3} - m(\cPqt))^2}{\sigma_{\mathrm{j}1\mathrm{j}2\mathrm{j}3}^2} + \frac{(M_{\mathrm{j}1\mathrm{j}2} - m(\PW))^2}{\sigma_{\mathrm{j}1\mathrm{j}2}^2},
\label{eq:ChiHad}
\end{equation}
where $M_{\mathrm{j}1\mathrm{j}2\mathrm{j}3}$ and
$M_{\mathrm{j}1\mathrm{j}2}$ are, respectively the invariant mass of the
three-jet system from the top quark and of the two jets posited to
originate from W boson decay;
$\sigma_{\mathrm{j}1\mathrm{j}2\mathrm{j}3}$ and
$\sigma_{\mathrm{j}1\mathrm{j}2}$ are the uncertainties of these
invariant masses. The $M_{\mathrm{j}1\mathrm{j}2\mathrm{j}3}$ value is
calculated after imposing a $M_{\mathrm{j}1\mathrm{j}2}=m(\PW)$
constraint by kinematic fit, while $M_{\mathrm{j}1\mathrm{j}2}$ is the
two-jet invariant mass before the fit. The jet assignments are made
according to the b tag information: $\mathrm{j}3$ must be tagged as a b
quark if there are at least two \bjets in the event, and $\mathrm{j}1$
and $\mathrm{j}2$ cannot be tagged unless there are at least three
\bjets in the event. This variable is used for the signal selection in
the tt decay mode.

\item \mtb, \mlb: Finally, to kinematically disentangle the signal from
the \ttbar background, we construct two new invariant-mass variables
that characterize the process where one \stp decays into 3 jets and \lsp
while the other decays into a b quark, lepton, neutrino, and \lsp.  In
the case of the bbWW decay mode and the tt decay mode where no on-shell
top quark is produced, i.e. $m(\stp)-m(\lsp)< m(\cPqt)$, the \mlb
distribution discriminates between \ttbar and signal. The quantity \mtb
is the invariant mass of the 3 jets among the 4 highest \pt jets which
are the most back-to-back (according to angular difference) to the
lepton. In the case of \ttbar background, \mtb reconstructs the mass of
the top quark having decayed into 3 jets, modulo the limitations of the
jet association. For the bbWW decay mode of the signal, it reconstructs
an invariant mass different from $m(\cPqt)$, as no top quark is present
in the final state. The quantity \mlb is defined as the invariant mass
of the lepton and the \bjet closest to it in $\Delta R$.

\end{itemize}

\begin{figure*}[htb]
\includegraphics[width=0.5\textwidth]{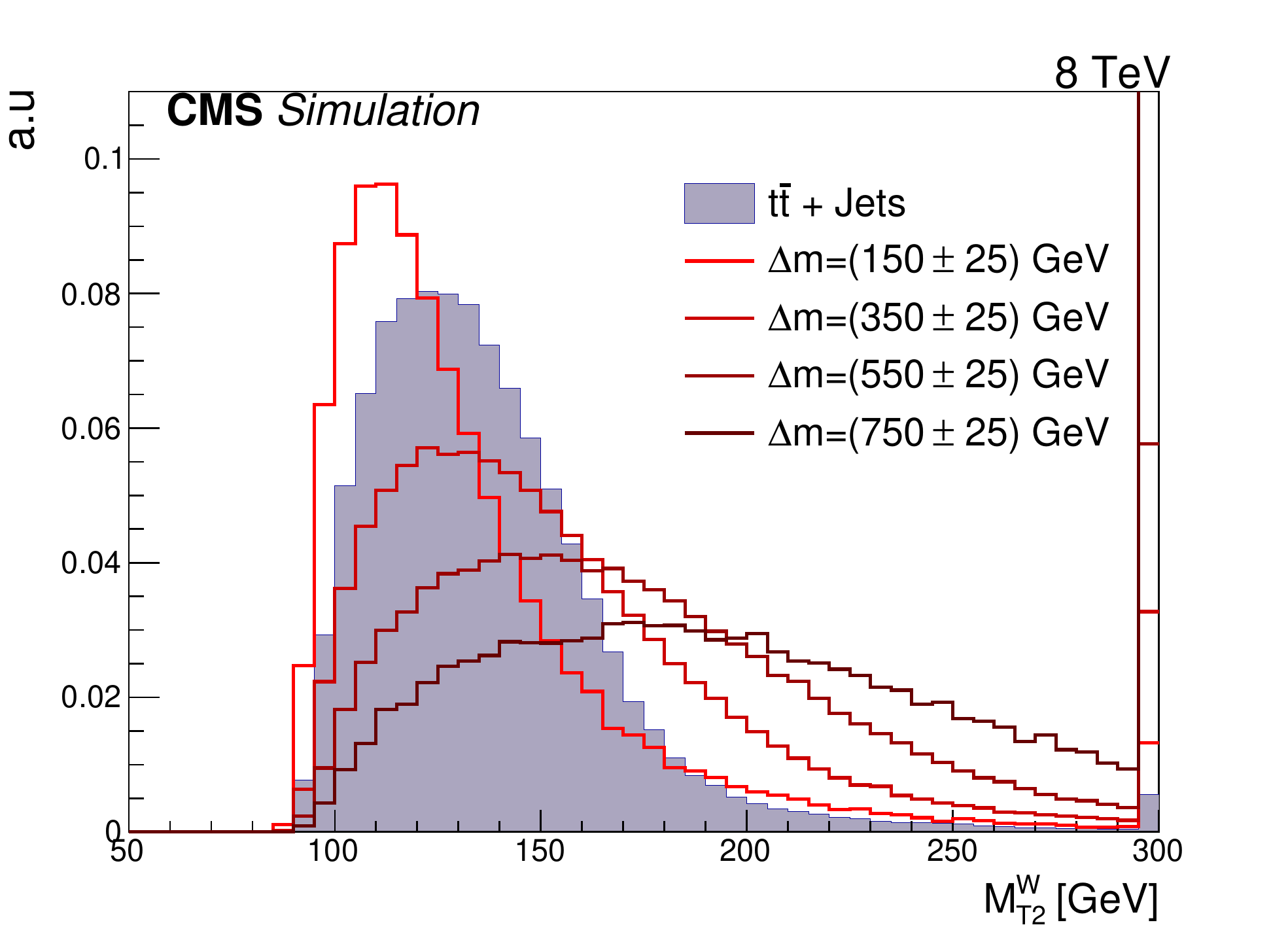}
\includegraphics[width=0.5\textwidth]{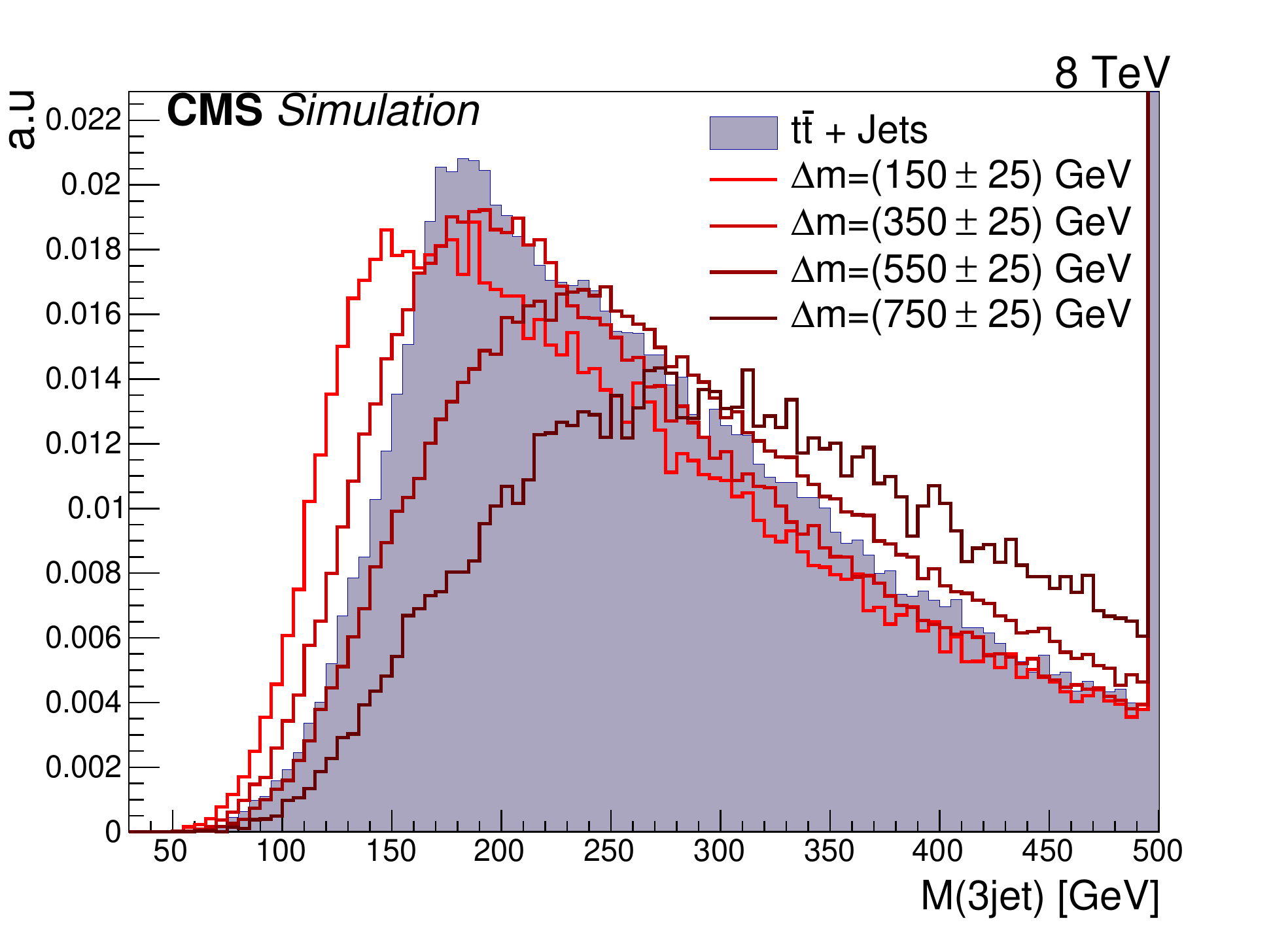}
\includegraphics[width=0.5\textwidth]{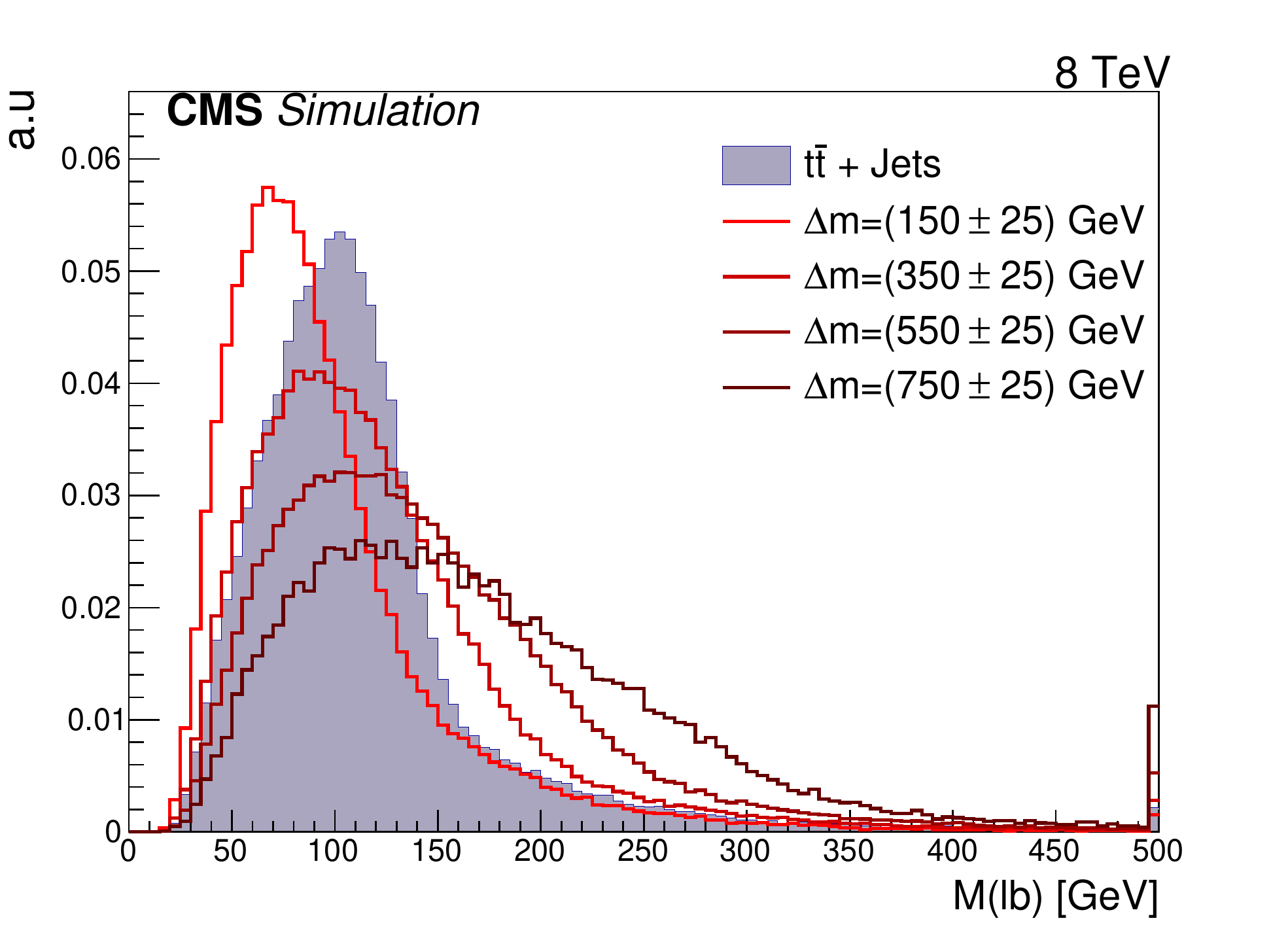}
\includegraphics[width=0.5\textwidth]{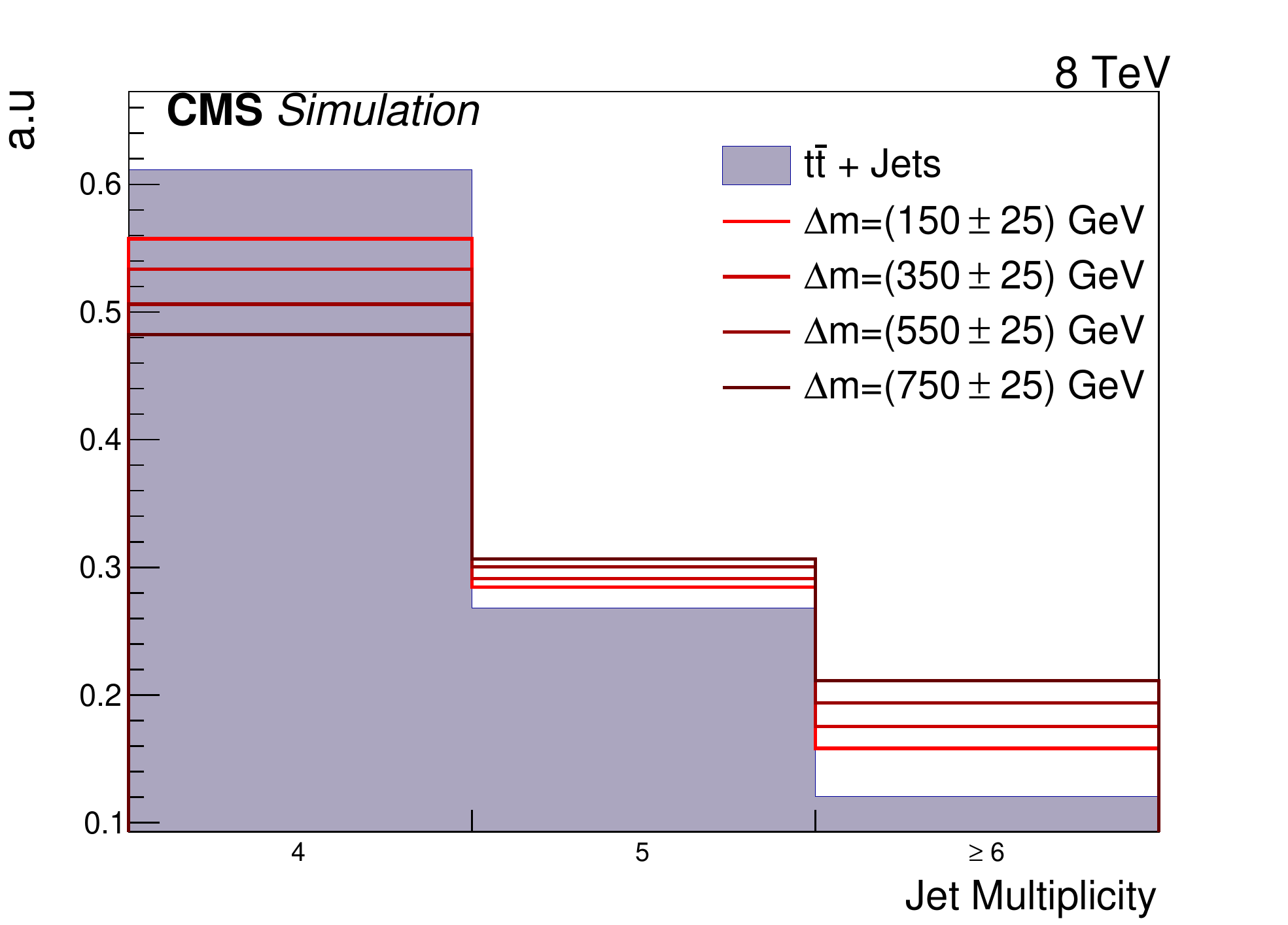}
 \caption{Distribution of some discriminating variables for the bbWW
($x=0.75$) decay mode at the preselection level, for the main \ttbar
background and benchmark signal mass points grouped in bands of constant
width $\dm = (150 \pm 25)$, $(350 \pm 25)$, $(550 \pm 25)$, and $(750
\pm 25)\GeV$. Distributions are normalized to the same area. From left
to right and from top to bottom: \mw, \mtb, \mlb and \njets.}
\label{fig:T2BW75}
\end{figure*}

\begin{figure}[!hbt]
\includegraphics[width=0.5\textwidth]{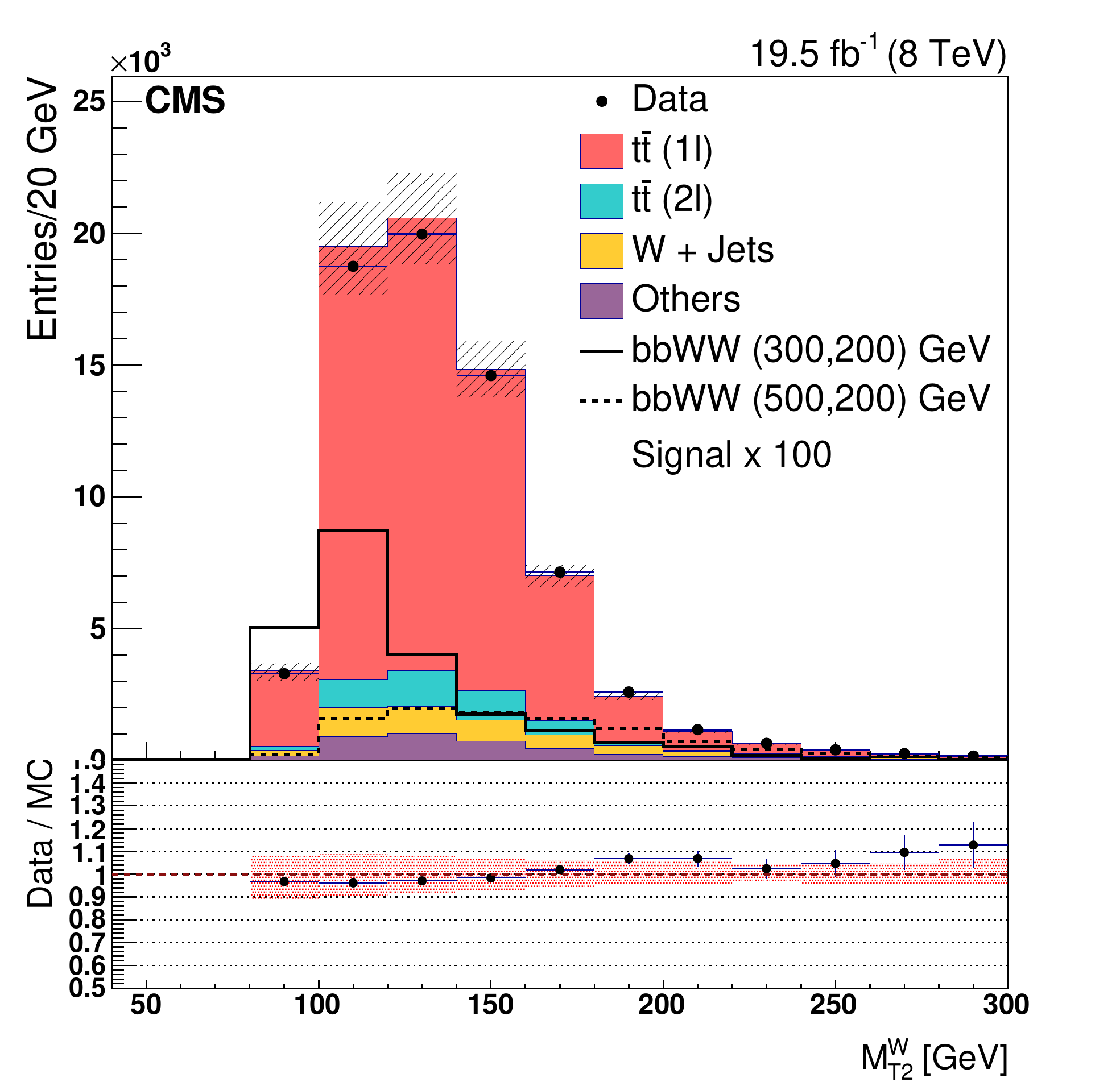}
\includegraphics[width=0.5\textwidth]{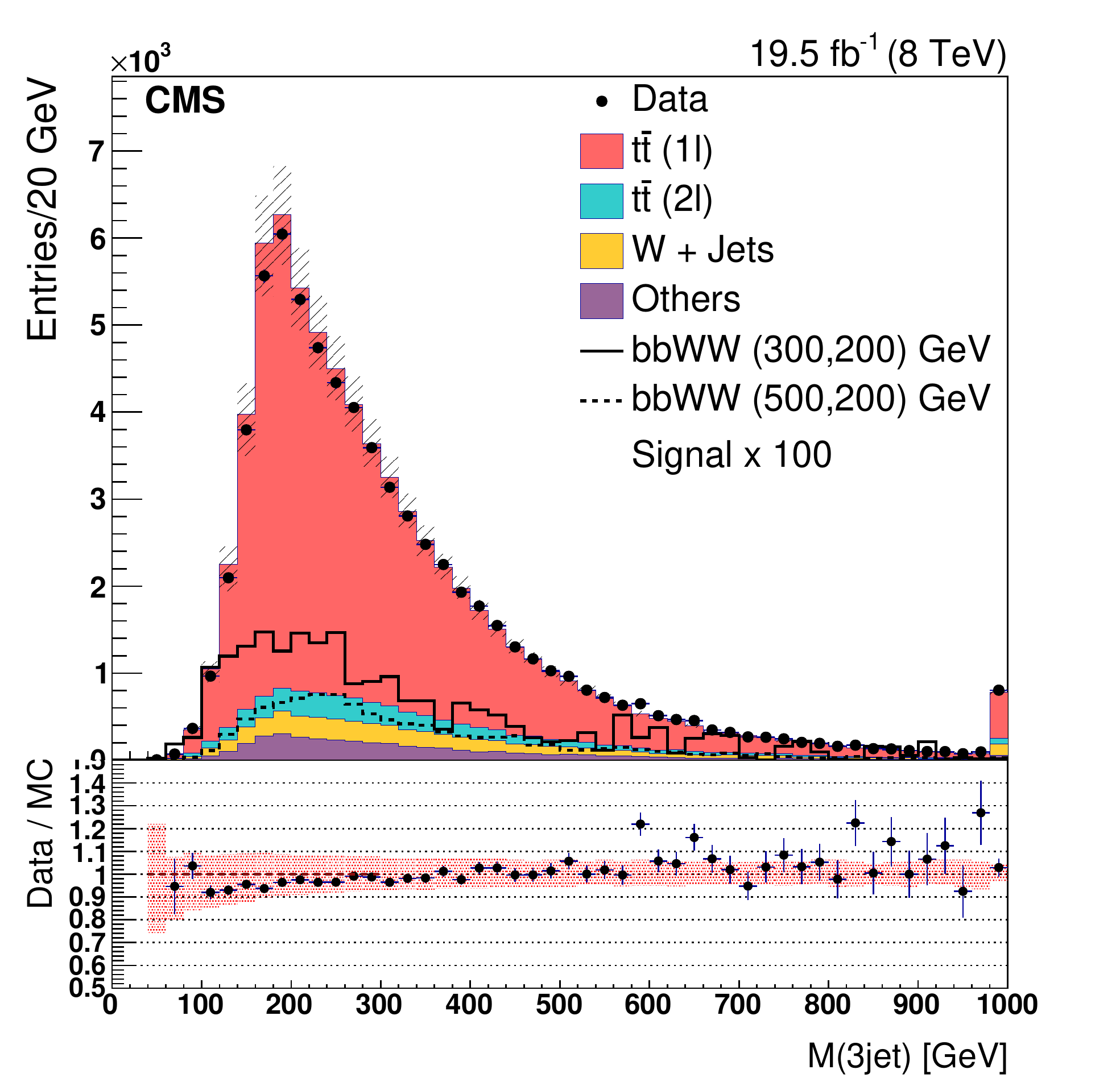}
\includegraphics[width=0.5\textwidth]{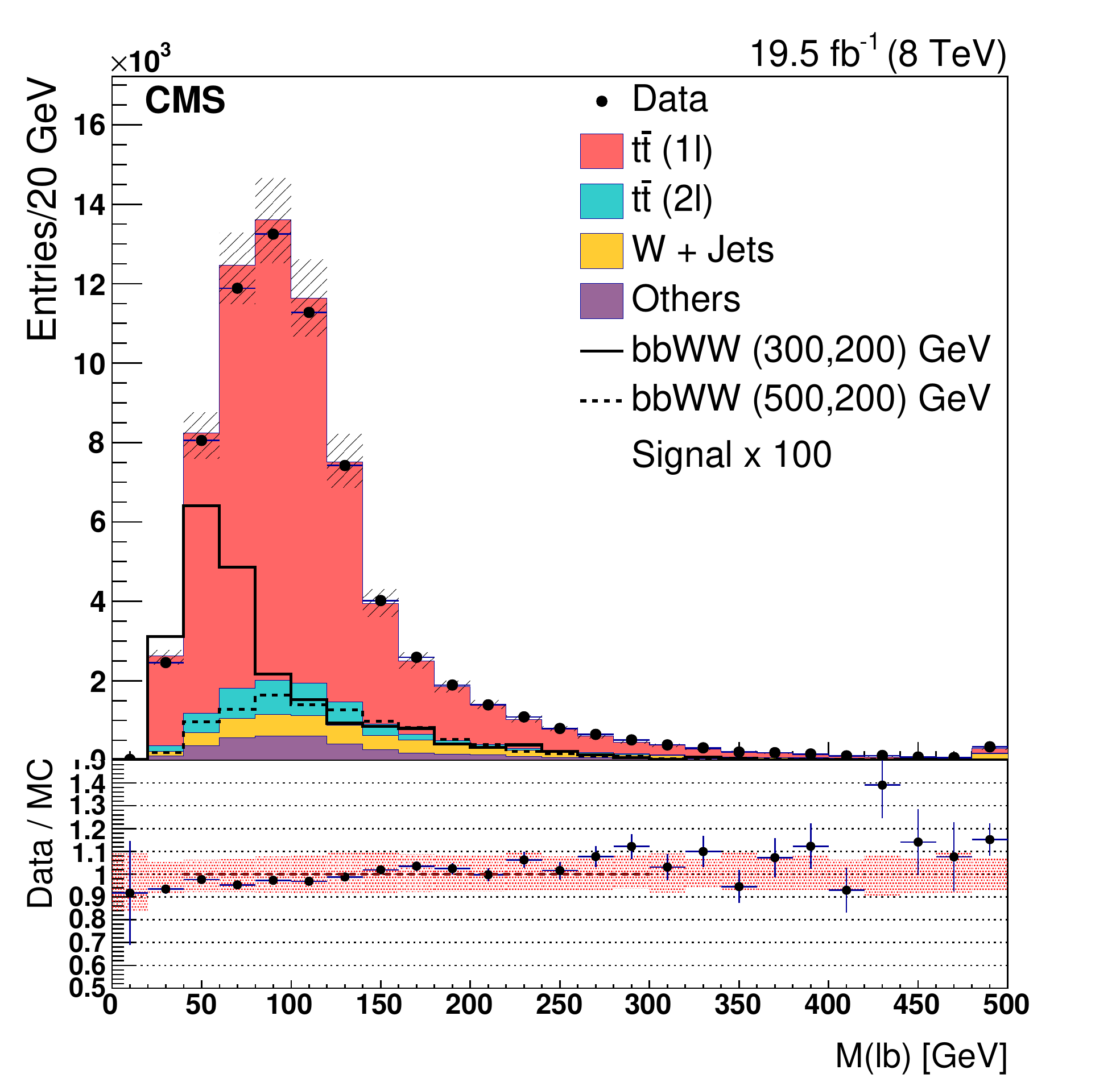}
\includegraphics[width=0.5\textwidth]{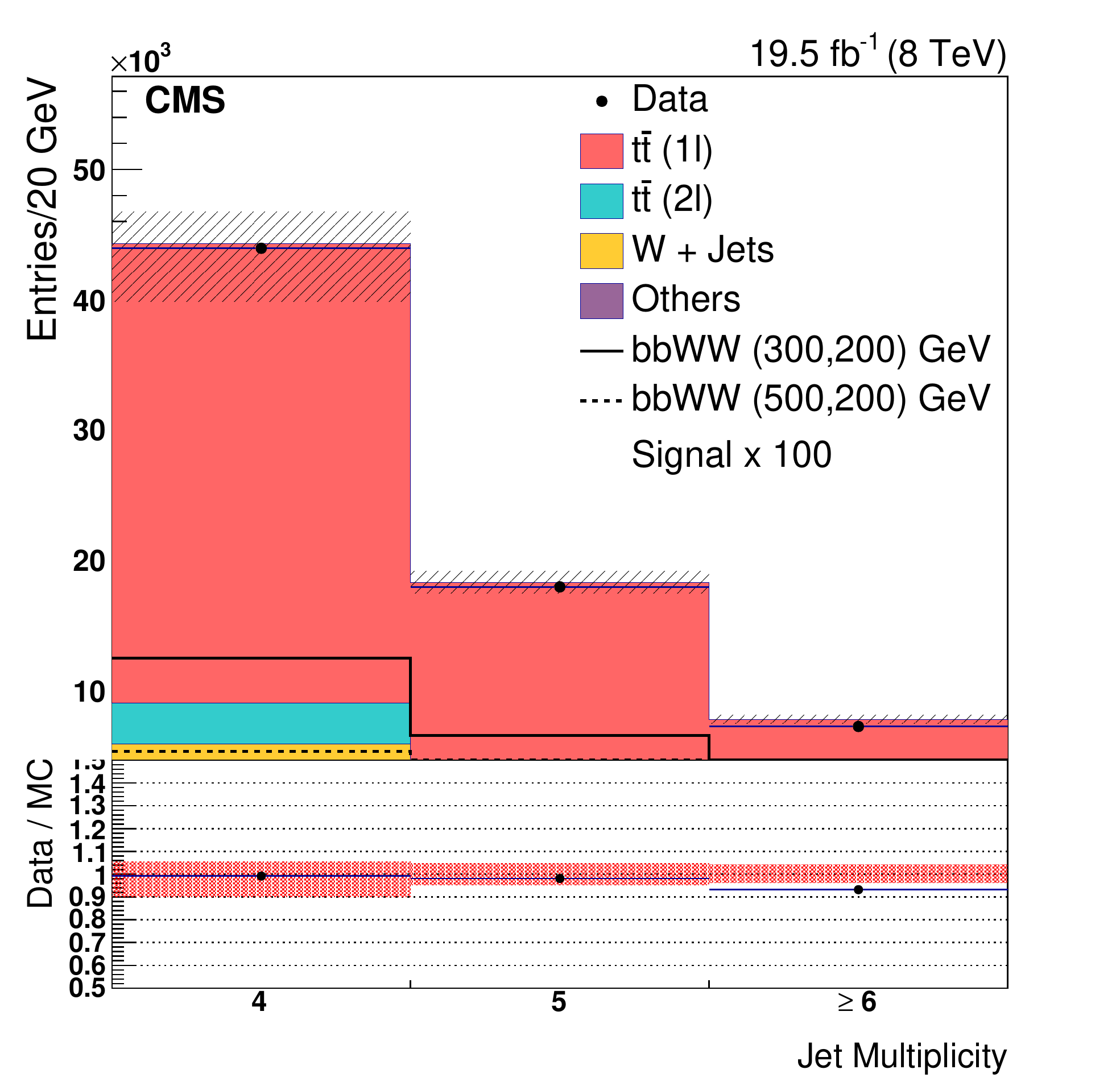}
 \caption{
Distributions of different variables in both data and simulation, for
both e and $\mu$ final states at the preselection level without the \mt requirement. From
left to right: \mw, \mtb, \mlb and \njets. The hatched region represents the
quadratic sum of statistical and JES simulation uncertainties. The lower
panel shows the ratio of data to total simulation background, with the
red band representing the uncertainties mentioned in the text. Two signal mass
points of the bbWW decay mode ($x=0.75$) are represented by open
histograms, dashed and solid, with their cross sections scaled by 100;
the two mass points \mmplane are (300, 200) and (500, 200)\GeV.}
\label{fig:DataMCpre1}
\end{figure}

Distributions of some of the variables used for the bbWW ($x=0.75$)
decay mode are illustrated in Fig.~\ref{fig:T2BW75}. The figure shows
the distributions for both \ttbar and signal samples; in the latter case
four different kinematic possibilities are illustrated, distinguished by
values of \dm:
\begin{equation}
\label{eq:DM}
\dm\equiv m(\stp)-m(\lsp).
\end{equation}
The figure shows clearly the evolution of the kinematic distributions as
the mass difference between the lightest top squark and the LSP is
varied. Differences in kinematic distributions may also be seen when
comparing the tt and bbWW signal decay modes, and when varying the
choice of $x$ ($x=0.25,$ 0.50, 0.75) in the bbWW decay mode. In
Fig.~\ref{fig:DataMCpre1} we show distributions of some discriminating
variables at the preselection level (but without the restriction on \mt)
for both e and $\mu$ final states in data and simulated events. The
figure shows good agreement between data and the total simulated
background, within the uncertainties of the simulated events, which
include the statistical uncertainty in the simulation samples
quadratically added to the systematic uncertainty in the jet energy
scale (JES).

As expected from the distributions shown in Fig.~\ref{fig:T2BW75}, different
selection variables will exhibit different degrees of discriminating power,
depending on the decay mode (tt or bbWW) and the relevant mass parameters
(\dm or $x$) of the signal. To find the most discriminating variables, we
test different sets of candidate BDT input variables, maximizing a figure of
merit that compares the expected signal yield to the quadratic sum of the
statistical and systematic uncertainties in the expected background yield.
To keep the selection tool
simple, a new variable is incorporated into the set of input variables only
if it leads to a substantial increase in the figure of merit.
The training of the BDT, together with this procedure for selecting variables,
is then carried out separately for the different decay modes tt and bbWW ($x=0.25$, 0.50,
0.75), and across six benchmark kinematic regions, defined as: $\dm =
(150\pm25),  (250\pm25),  (350\pm25),  (450\pm25),  (550\pm25),
\text{ and } (650\pm25)$\GeV. This partitioning
allows us to take into account the evolution of the signal
kinematics across the \mmplane plane.
The different BDT trainings are numbered from 1 to 6 to reflect
the \dm regions in which they are trained.

\begin{figure*}
\begin{tabular}{ll}
\includegraphics[scale=0.44]{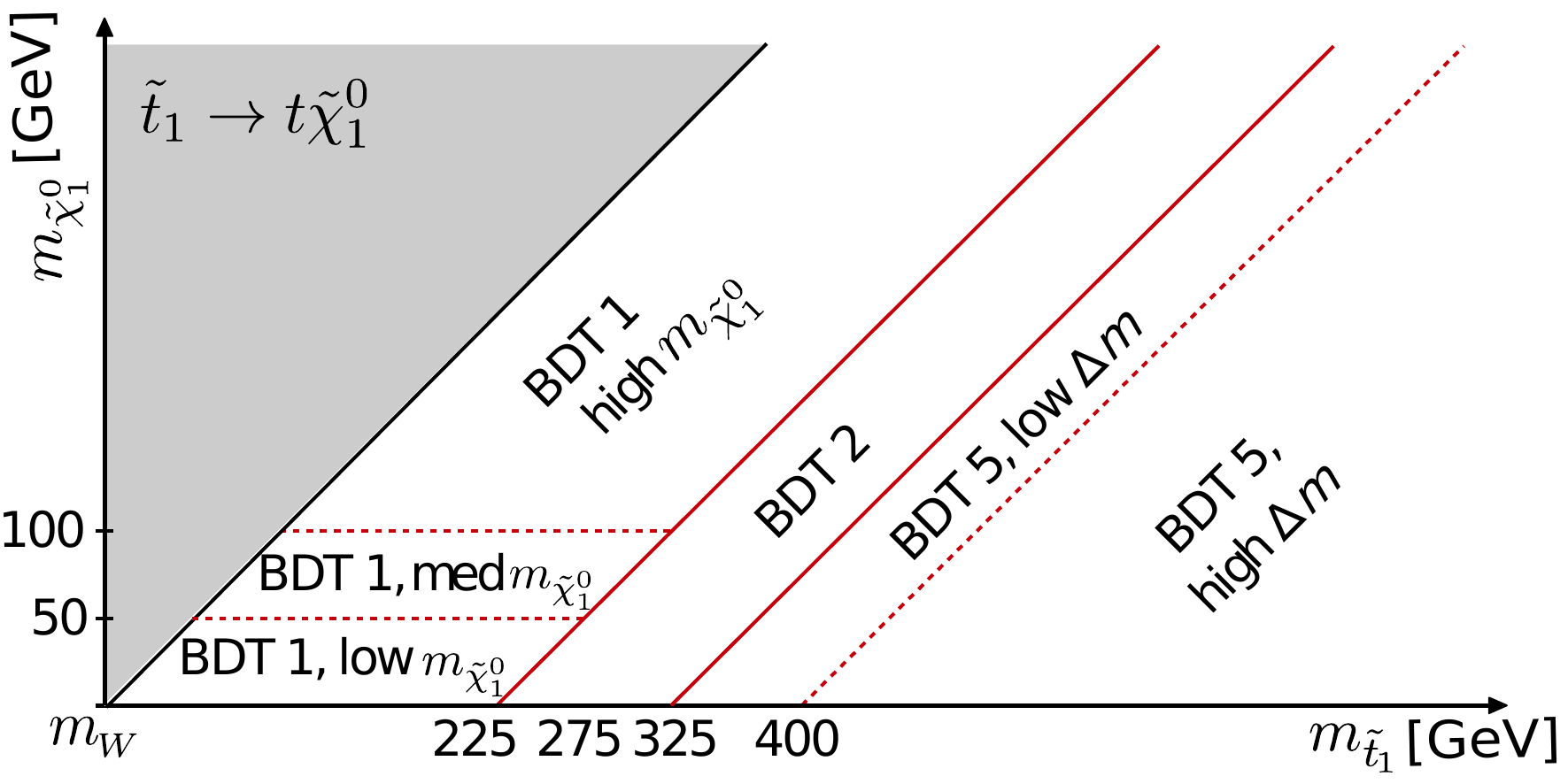}
\includegraphics[scale=0.44]{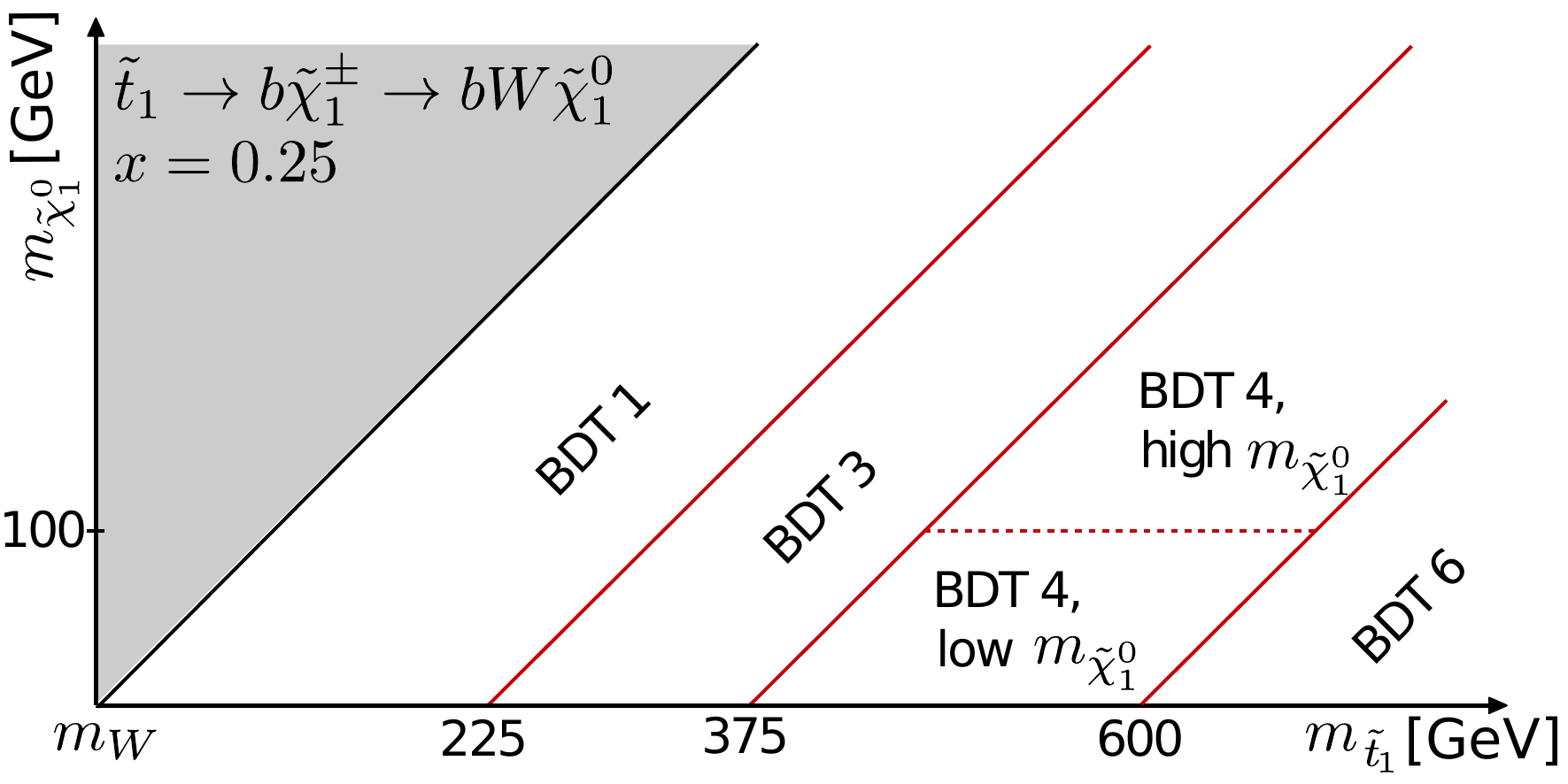} \\
\includegraphics[scale=0.44]{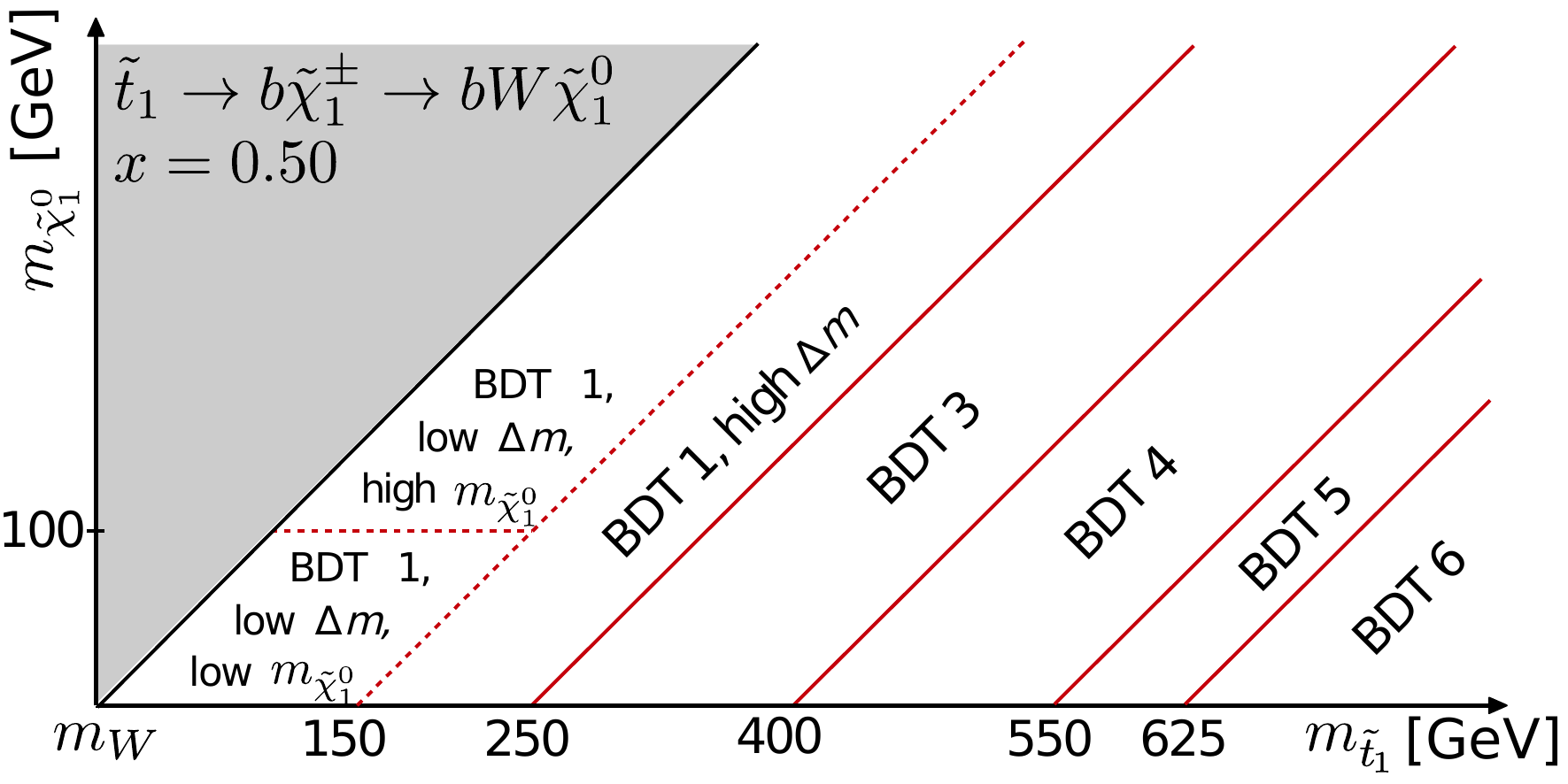}
\includegraphics[scale=0.44]{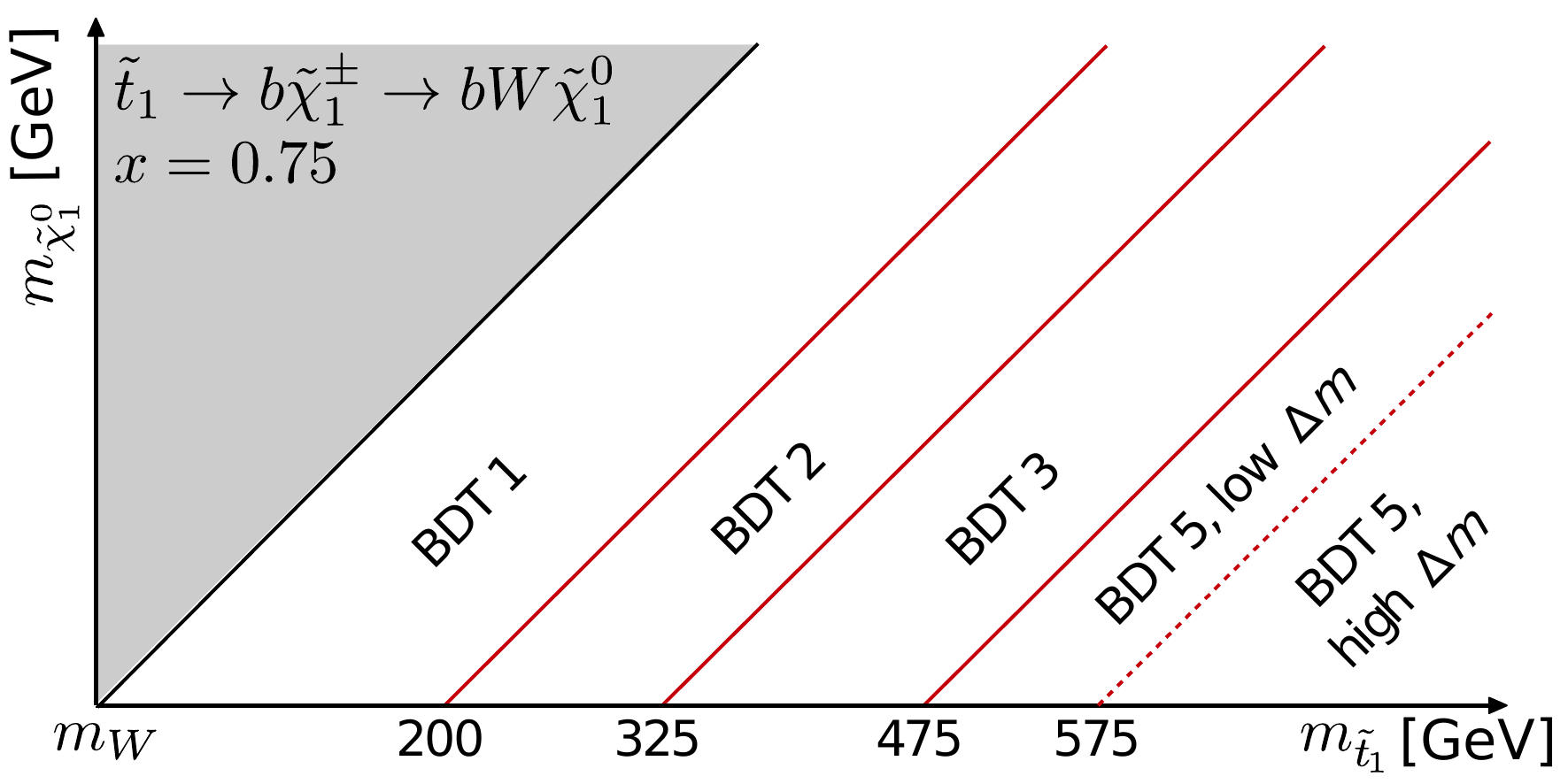}
\end{tabular}
\caption{Signal regions (SRs) defined as functions of the chosen BDT
trainings in the \mmplane plane for tt (top left), bbWW $x=0.25$ (top
right), 0.50 (bottom left), and 0.75 (bottom right) decay modes. The
SRs are delimited by continuous red lines, and the final selections
within the different SRs are delimited by dashed red lines. The
attributes ``low / high m(\lsp)'' and ``low / high \dm'' indicate
that in these regions different thresholds are applied for the same
BDT training.}
\label{fig:BestBDT}
\end{figure*}

The final sets of variables retained as input to the BDT are reported in
Table~\ref{tab:bdtVar}. Having been chosen with a quantitative
assessment of the discriminating power of each variable, these represent
the most reduced, while effective, sets of input variables to the BDT,
for each decay mode and kinematic region. This represents a new feature
of this search compared to Ref.~\cite{MtPub}, where the BDT was trained
with the same set of variables across different kinematic regions. Once
the input variables to the BDT are determined, different BDTs are
trained in each of the benchmark kinematic regions to build selection
tools adapted to a kinematically varying signal.  The simulation samples
used for finding the best set of variables and training the BDT are
statistically independent. This procedure is done for the tt and bbWW
(different $x$ values) decay modes. Using a more systematic approach for
the definition of signal regions (SRs) than in Ref.~\cite{MtPub}, we
first consider which training is the best performing one in the
\mmplane plane. We observe that some BDTs are the best over a very
limited part of the \mmplane plane, so to simplify the final selection
we retain BDT trainings that are observed to be the best performing over
a large portion of the mass plane. The resulting SRs, defined as the
chosen BDT training in the \mmplane plane, are shown for all considered
decay modes in Fig.~\ref{fig:BestBDT}.

\begin{table}[!htbp]
\centering
\topcaption{Final selection variables chosen as input for the BDT training, as
functions of the decay modes bbWW and tt, and kinematic regions.
Column headings \dr\ and \df\ refer to \drLB and \dfMJ.}
\resizebox{\textwidth}{!}{
\begin{tabular}{@{\extracolsep{\fill}}l|*{13}{c}}
\hline
                          &\met& \ptl&  \mw& \njets& \ptj& \ptb&  \HT& \HTfrac& \dr&  \df& \ChiHad& \mlb& \mtb~   \\
\hline
{tt:}                 &    &     &     &       &     &     &     &        &    &     &        &     &        \\
$\Delta m < m(\cPqt)$       &\ch & \ch &     &  \ch  &     &  \ch&     &     \ch&    &  \ch&        &  \ch&        \\
$\Delta m \geq m(\cPqt)$    &\ch & \ch &  \ch&  \ch  & \ch &     &     &     \ch& \ch&  \ch&  \ch   &     &        \\
\hline   %
{bbWW:}               &    &     &     &       &     &     &     &        &    &     &        &     &        \\
$x=0.25$, 0.50            &\ch & \ch &  \ch&  \ch  &     &  \ch&     &        & \ch&  \ch&        &  \ch& \ch    \\
$x=0.75$                  &\ch & \ch &  \ch&  \ch  & \ch &     &  \ch&        &    &  \ch&        &  \ch& \ch    \\
\hline
\end{tabular}
}
\label{tab:bdtVar}
\end{table}

\begin{figure}[!hbt]
\centering
\includegraphics[width=0.49\textwidth]{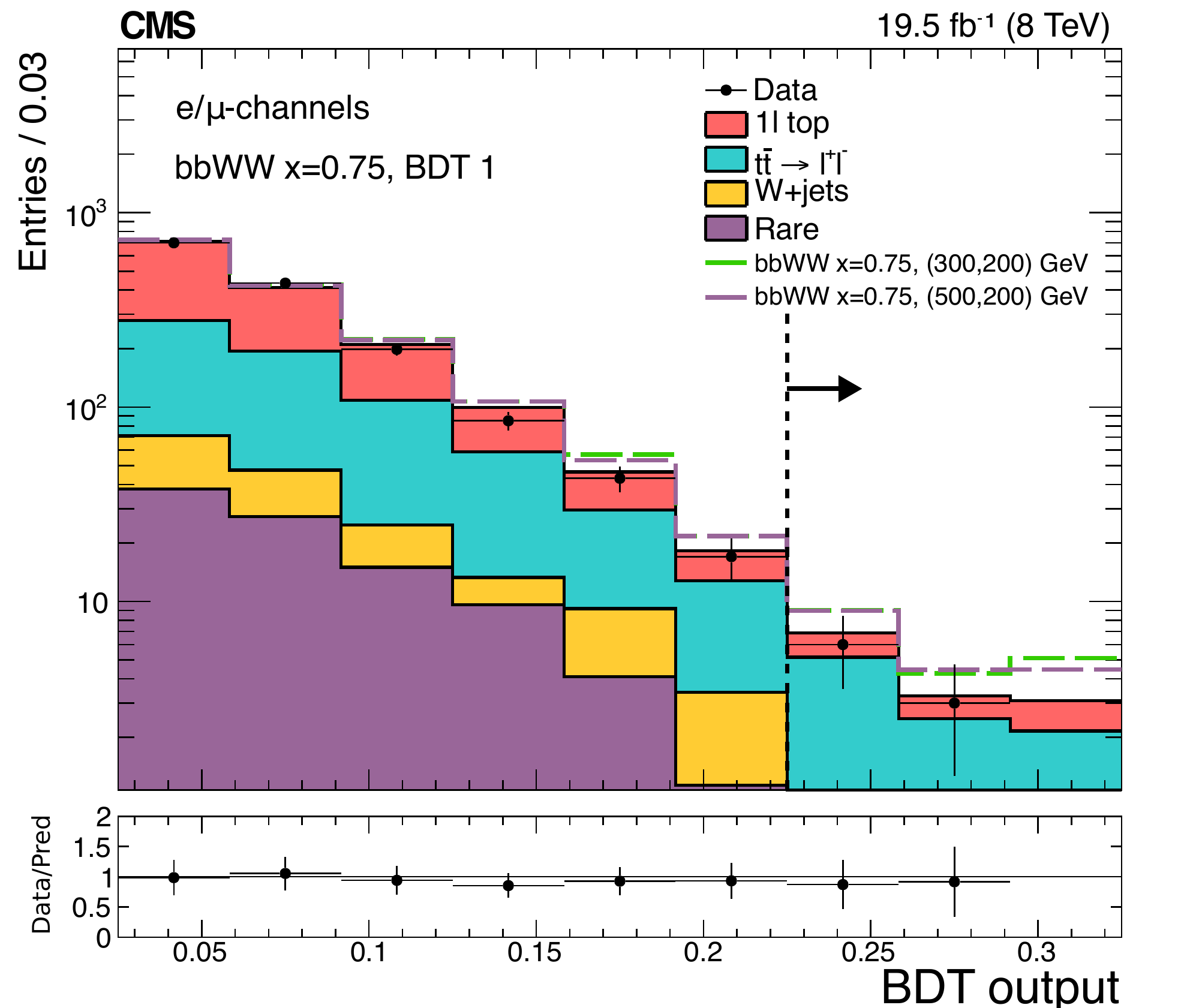}
\includegraphics[width=0.49\textwidth]{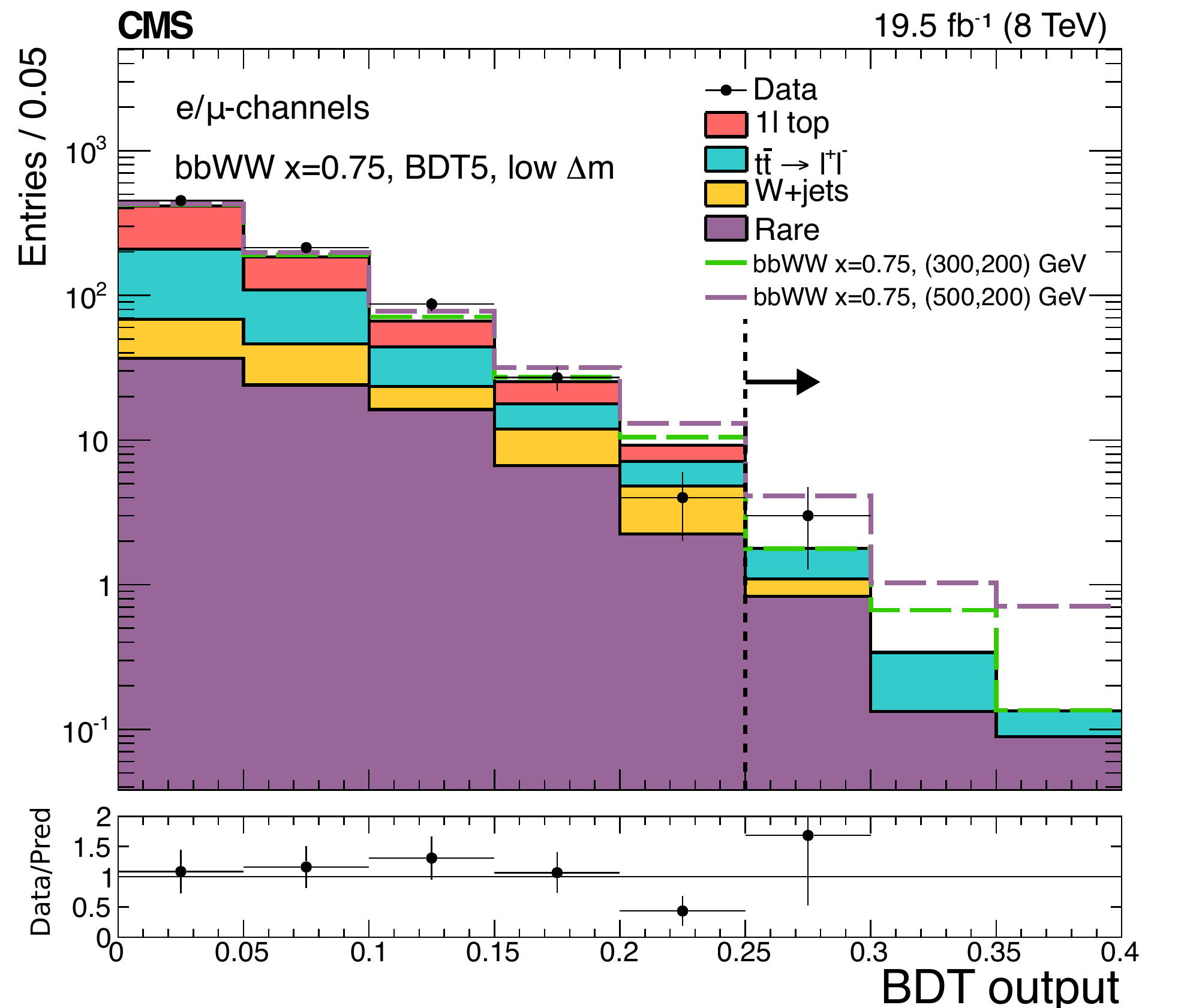}
\caption{
\label{fig:BDToutDD075}
The BDT output distributions of the bbWW $(x=0.75)$ decay mode in both
final states at the preselection level for data and predicted
background, with $\mathrm{BDT1} >0.025$ (left) and $\mathrm{BDT5}>0$
(right). Two representative signal mass points are shown: $\mmplane =
(300,200)$ and (500,200)\GeV. In each panel the final selection is
indicated by the vertical black dashed line. The normalization and \mt
correction (see Section~\ref{s:0btag}), computed in the tail of the BDT
output, i.e. to the right of the dashed line, are here propagated to the
full distribution. The uncertainties are statistical. The plots on the
bottom represent the ratio of Data over the predicted background, where
we quadratically add statistical uncertainties with the uncertainties on
the scale factors.}
\end{figure}

With these SRs determined, the final selection is made by applying a
minimum threshold to each BDT output as shown in
Fig.~\ref{fig:BDToutDD075}. The thresholds are determined by minimizing
the expected upper limit cross section ($\sigma_{95}^\text{exp}$) obtained
from events remaining above the threshold, taking into account the
predicted background (Section~\ref{s:1lepBckg}). The final BDT trainings
and selections are reported in
Fig.~\ref{fig:BestBDT} for all decay modes; within some SRs, the same
BDT training is used with different threshold values, thus leading to different selections.
On average the BDT
selection suppresses the SM background by a factor ${\sim}10^3$ while
reducing the signal only by a factor $\sim$10; the performance improves
monotonically with increasing \dm.

\subsection{Background estimation}
\label{s:1lepBckg}

The SM background processes in the single-lepton search can be divided
into four categories. At preselection, the dominant contribution ($\sim$
66\% of the total) is the \ttbar production with one lepton; we include
single top-quark production in this category and call the combination
the ``\ttl'' component. The second most significant background (23\%)
comes from \ttbar\ events with two leptons, where one lepton escapes
detection; we will call this the ``\ttll'' component. The third
background (7\%) is the production of \PW\ in association with jets,
which we will denote ``\wjets''. Other backgrounds are labeled as
``rare''. We use data to estimate the event yields of the first three
categories, starting with distributions obtained from simulation, and
normalizing these with scale factors ($SF$) determined in control
regions.
The background is estimated using the formulae:
\begin{equation}\begin{aligned}
\label{eq:form1ltop}
N_\text{tail}(\ttl)                      & =   \SFpost\,  N^\text{MC}_\text{tail}(\ttl)        \, SFR_{1 \ell}, \\
N_\text{tail}(\cPqt\cPaqt \to \ell\ell)   & =   \SFpre\, N^\text{MC}_\text{tail}(\cPqt\cPaqt \to \ell\ell),     \\
N_\text{tail}(\PW\text{+jets})                   & =   \SFpost  \,  N^\text{MC}_\text{tail}(\PW\text{+jets})     \, SFR_{\PW}.
\end{aligned}\end{equation}
The subscript \textit{tail} refers to the region $\mt > 100$\GeV. The
simulation yields at the final selection level
($N^\text{MC}_\text{tail}$) are corrected by normalization scale factors
$\SFpre$ and $\SFpost$ (defined in Eq.~\eqref{eq:sf1}
and~\eqref{eq:sf2}), determined in the \mt peak region $50< \mt
<80$\GeV. The additional scale factor ratios, denoted $SFR_{1 \ell}$ and
$SFR_\text{W}$, are used to correct the tail of the \mt distribution,
and are determined using a control region with zero \bjets. The
procedure accounts for the possibility of signal contamination in the
different control regions. At the final selection level the \ttll\
process represents an approximately constant proportion of the total
background at $\sim$60\%, while the \ttl\ and \wjets processes have
varying proportions across the different selections within the remaining
$\sim$40\%. Signal contamination is important only at low \dm, where it
alters the background determination by up to 25\%.

\subsubsection{Normalization in the \mt peak}
\label{s:peak}

The scale factors $\SFpre$ and $\SFpost$ are estimated to correct for the
normalization in the \mt peak region and after the final selection on the
output of the BDT.
To calculate $\SFpost$ we further require the second lepton veto,
while $\SFpre$ is obtained without this veto. $\SFpre$ fixes the \ttll\
background normalization, while $\SFpost$ sets the \ttl\ and \wjets
background normalizations. The scale factors are computed as follows:
\begin{eqnarray}
        \SFpre & = & \left( \frac{N(\text{data}) - N^{\text{MC}}(\text{rare}) - N^{\text{MC}}(\text{signal})}{N^{\text{MC}}(\ttl) + N^{\text{MC}}(\cPqt\cPaqt \to \ell \ell) + N^{\text{MC}}(\PW\text{+jets})} \right),
    \label{eq:sf1} \\
        \SFpost & = & \left( \frac{N(\text{data}) - N^{\text{MC}}(\text{rare})  - N^{\text{MC}}(\text{signal}) - \SFpre \, N^{\text{MC}}(\cPqt\cPaqt \to \ell \ell)}{N^{\text{MC}}(\ttl) + N^{\text{MC}}(\PW\text{+jets})} \right).
    \label{eq:sf2}
\end{eqnarray}
The inclusion of the $N^{\text{MC}}$(\text{signal}) term accounts for
possible signal contamination. At preselection we have: $\SFpre$ = ($1.06
\pm0.01$) and $\SFpost$ = ($1.05 \pm 0.01$). At the final selection
level, the deviation of these scale factors from unity is always within
10$\%$.

\subsubsection{Correction for the tail in the \mt distribution}
\label{s:0btag}

\begin{figure}[!hbt]
  \centering
    \includegraphics[width=0.45\textwidth]{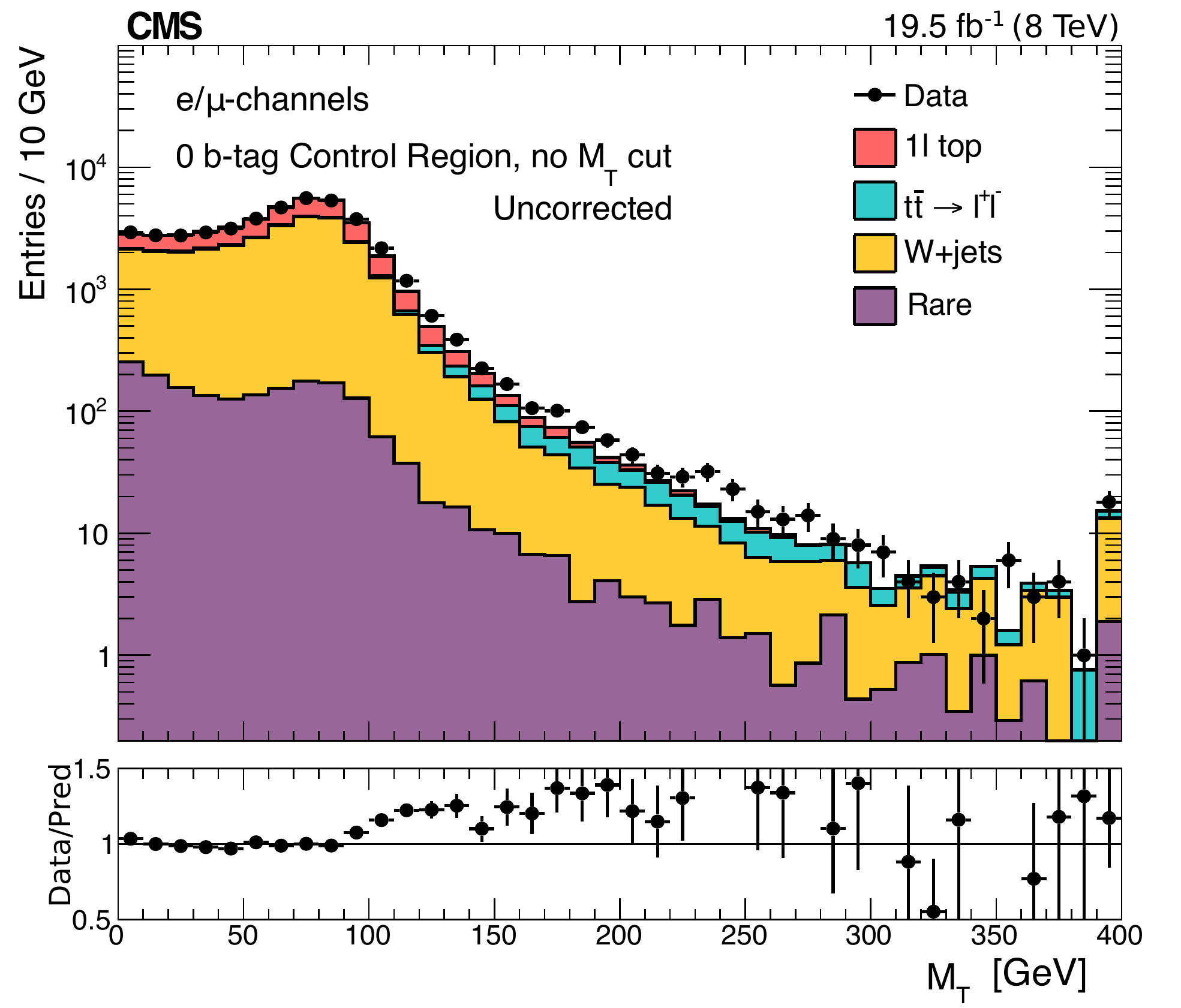}
    \includegraphics[width=0.45\textwidth]{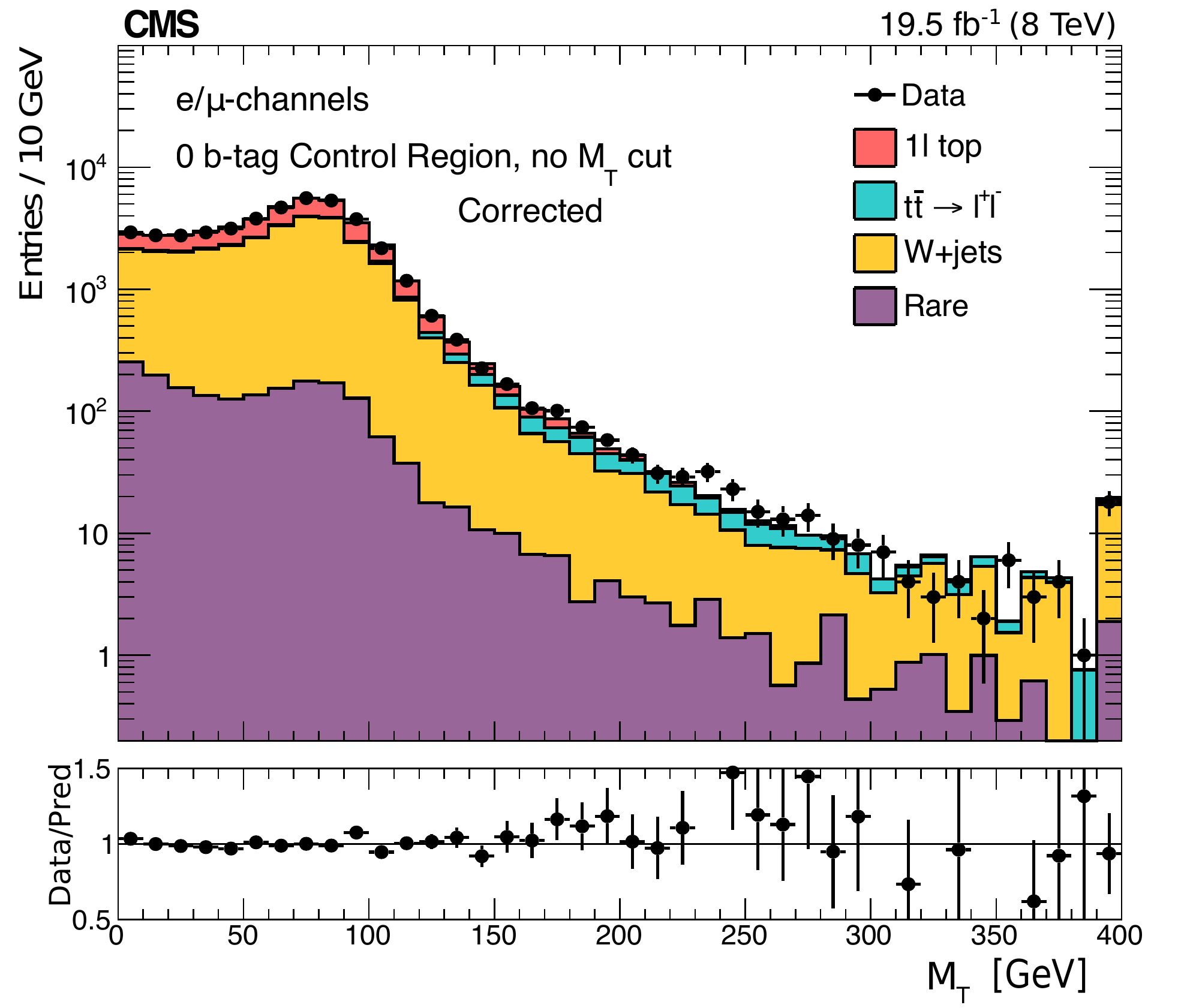}
\caption{
\label{fig:cr_0btag_full_mt}
Full \mt distribution in the control region with zero \bjets, without
any extra signal selection. Left: without the tail correction factors
applied; right: with $SFR_{\PW}$ and $SFR_{\text{1$\ell$}}$ corrections
applied. The plots on the bottom represent the ratio of Data over the
predicted background.}
\end{figure}

To study the tail of the \mt distribution for different backgrounds, we
enrich the data with the \wjets contribution by inverting the b-tagging
criterion of the preselection. The left plot of
Fig.~\ref{fig:cr_0btag_full_mt} compares the data with background
simulation, and shows some disagreement between the two for
$\mt>100$\GeV. To correct this, we follow an approach based on template
fits, which allows us to extract different correction factors for the
\ttl\ and the \wjets\ backgrounds, rather than assuming them to be equal
as in Ref.~\cite{MtPub}.

\begin{figure}[!hbt]    \centering
           \includegraphics[width=0.45\textwidth]{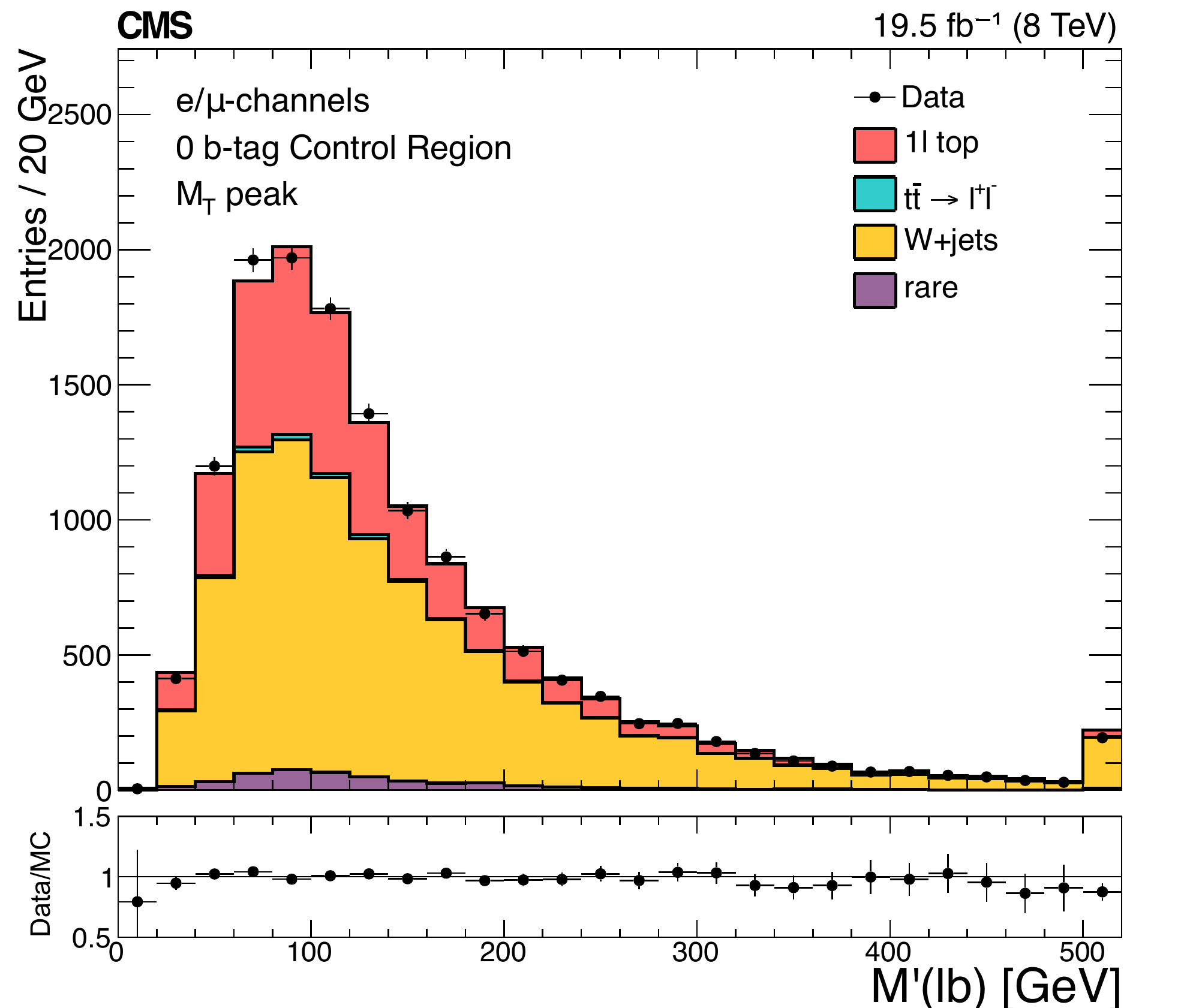}
           \includegraphics[width=0.45\textwidth]{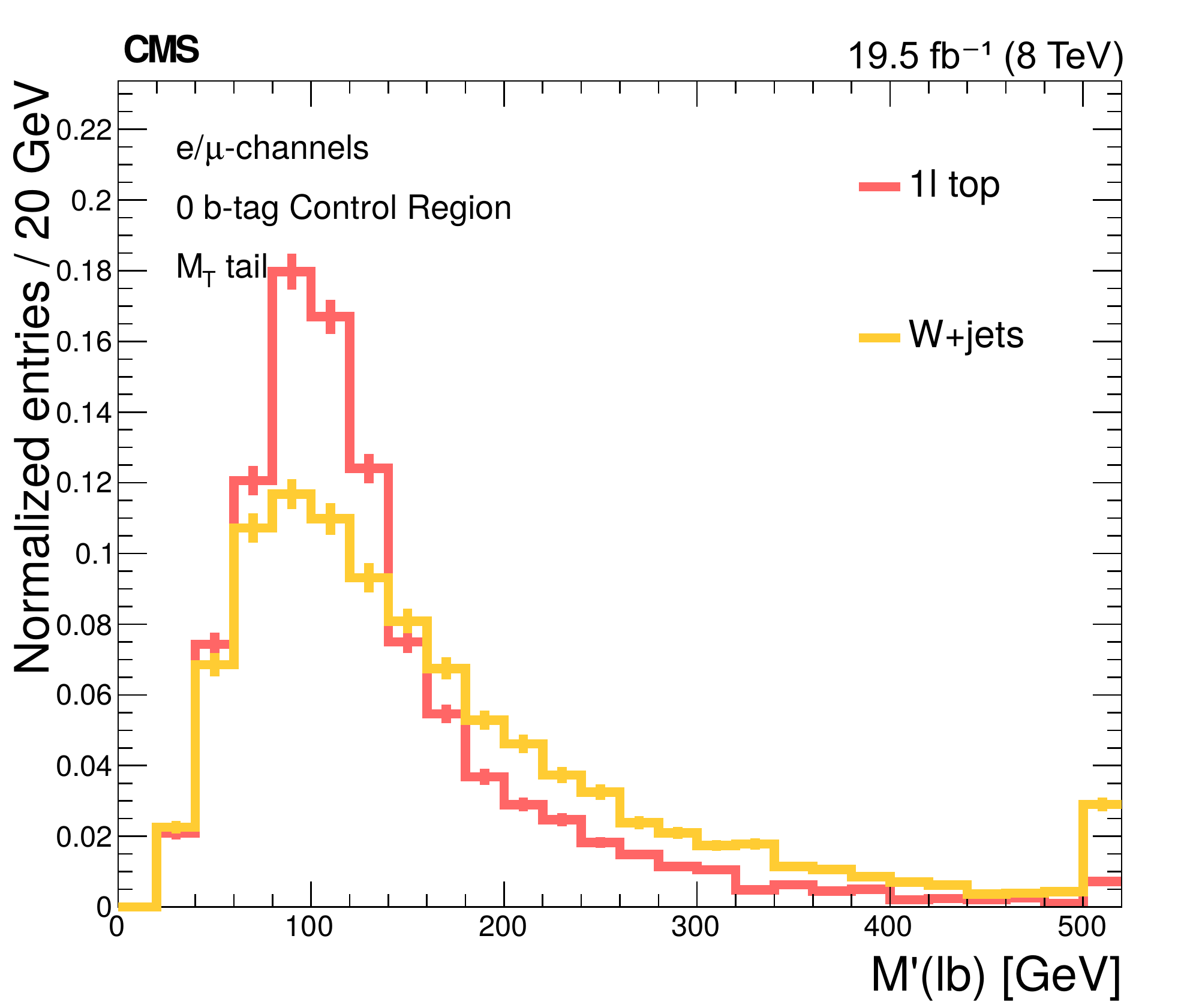}
\caption{
\label{fig:template_closure}
Left: Comparison of data and simulation in the \MTlb distributions
for events with $50<\mt<80$\GeV and zero \bjets. Right: Shape comparison
between \ttl\ and \wjets for $\mt>100$\GeV.}
\end{figure}

The template fit is performed using the invariant mass of the lepton and
the jet with the highest b-tag discriminator. This variable, \MTlb, is
well modeled by the background simulation (see
Fig.~\ref{fig:template_closure}, left) and exhibits discriminating power
between \wjets\ and \ttl\ (Fig.~\ref{fig:template_closure}, right). The
contributions of the \ttdl\ background, the rare backgrounds, and the
signal, are taken from simulation and their normalizations are
constrained within a 20\% uncertainty during the template fit. The
normalizations of the \ttl\ and \wjets\ backgrounds are free parameters
expressed in terms of scale factors $SF$. The fit is performed in a
control region with zero b-tag jets, in two separate regions of the \mt
distribution: the peak defined by $50<\mt<80$\GeV, and the tail defined
by $\mt>100$\GeV. We then extract the normalization independent ratios
$SFR = SF_\mathrm{tail}/SF_\mathrm{peak}$ for \ttl\ and for \wjets.
Without any BDT signal selection and for a case of negligible signal
contamination, the fit yields:
	$SFR_{\ttl} = (1.04 \pm 0.16$) and
	$SFR_{\PW} = (1.33 \pm 0.10$).
The right plot of Fig.~\ref{fig:cr_0btag_full_mt} confirms the
effectiveness of this correction.

Due to the low yields after the final selections, we loosen the
requirements on the output of the BDT to keep 25\% of the total yield
when we extract the $SFR$ values. The $SFR$ ratios obtained for the
different signal regions within a given decay mode (tt or bbWW) agree
well with each other. We therefore set the final $SFR$ factor for each
decay mode to the average over the signal regions for that mode. The
resulting $SFR$ values for the tt and bbWW decay modes differ from one
another, and also vary across the \mmplane mass plane:
$SFR_{\text{1$\ell$}}$ increases from 1.0 to 1.4 with increasing top
squark mass, while $SFR_{\PW}$ is stable around a mean value $\sim$1.2
everywhere. In addition to the extraction of tail correction factors, we
check in the control region with zero \bjets that the distributions of
all input variables in data are well described by the predicted
background.

\subsection{Systematic uncertainties}
\label{s:1lepSys}

The sensitivity of this search is limited by uncertainties in both the
background prediction and the acceptance and efficiency of the signal at
the mass points under consideration. The uncertainties are listed below.

\subsubsection{Background}
\label{s:1lepBSys}

For systematic uncertainties affecting the predicted background:
\begin{itemize}

\item We study the impact of limited simulation statistics, generator scale
variations, and JES uncertainty in the template fit method in the
control region with zero \bjets and no BDT selection. This leads to a
global absolute uncertainty of 0.6 in  $SFR_{1 \ell}$ and  0.4 in
$SFR_{\PW}$.

\item The goodness of the \ttll\ background modeling is checked in two
different control regions. The first uses events with exactly two
leptons in the final state and a lower jet multiplicity ($\njets\ge2$)
than that employed in the preselection; the second uses events with
exactly one lepton, and an isolated track or $\tauh$ candidate. The
simulation prediction is compared with data in the \mt tail region of
these control regions for each BDT selection. The comparison shows
overall agreement and deviations are used to derive a relative
systematic uncertainty, ranging from 20 to 80\% depending on the
selection.

\item We check the modeling of the \njets distribution in the \ttbar
background with a control region defined to have exactly two leptons and
no requirement on \mt. The data/simulation scale factors are observed to
be compatible with unity; therefore, no correction factor is used, but
the deviations from unity are taken as systematic uncertainty. This
leads to a flat 2\% uncertainty, used for all the BDT selections.

\item A 6\% uncertainty for the modeling of the isolated track veto is
applied to the fraction of \ttbar dilepton background events that have a
second $\Pe$/$\mu$ or a one-prong $\tauh$ decay in the acceptance. A 7\%
uncertainty for the modeling of the hadronic $\tau$ veto is only applied
to the fraction of \ttbar dilepton background events that have a $\tauh$
in the acceptance.

\item The $\SFpre$ and $\SFpost$ normalization factors are varied within
their statistical uncertainties and the variations are propagated as
systematic uncertainties to the \mt peak regions.

\item The statistical uncertainties in the simulation background samples
are propagated to the systematic uncertainties in the backgrounds.

\item The cross section of \wjets and rare backgrounds are
conservatively varied by 50\%, affecting the prediction of other
background processes through $\SFpre$ and $\SFpost$ (see equations of
Section~\ref{s:1lepBckg}); the cross section of the \ttbar process is
varied by 10\%.

\end{itemize}
Table \ref{tab:syst_summary} gives a summary of the relative systematic
uncertainties in the predicted total background yield at the
preselection level, as well as their range of variation over the
different top squark decay modes and BDT selections.

\begin{table}[!ht]
\centering
\topcaption{Summary of the relative systematic uncertainties in the total background, at
the preselection level, and the range of variation over the BDT selections.}
\newcolumntype{.}{D{.}{.}{6}}
\newcolumntype{X}{D{,}{\text{---}}{8}}
\begin{tabular}{l.X}
\hline
Source & \multicolumn{1}{c}{Uncertainty (\%)}  & \multicolumn{1}{c}{Uncertainty (\%) range}  \\
& \multicolumn{1}{c}{at preselection}            &  \multicolumn{1}{c}{over BDT selections} \\
\hline
\text{$SFR_{1 \ell}$ uncertainty}                           & 16.4     & 0,24    \\
\text{$SFR_{\PW}$ uncertainty}                               & 1.4      & 0,5    \\
\text{Modeling of \mt tail in $\cPqt\cPaqt \to \ell \ell$} & 1.6      & 7,39    \\
\text{Modeling of \njets in $\cPqt\cPaqt$}                          & 1.1      & 1,4    \\
\text{Modeling of the 2$^\mathrm{nd}$ lepton veto}                     & 1.2      & 1,4    \\
\text{Normalization in $\mt$ peak (data \& MC stat)}            & 0.7      & 3,37    \\
\text{Simulation statistics in SR}                                      & 0.4      & 3,38    \\
\text{Cross section uncertainties}                              & 2.0      & 4,34    \\
\hline
\text{Total}                                                    & 16.8     & 23,58    \\
\hline
\end{tabular}
\label{tab:syst_summary}
\end{table}

\subsubsection{Signal}
\label{s:1lepSSys}

The statistical uncertainties in the signal samples are taken into
account.  The integrated luminosity is known~\cite{CMS:2013gfa} to a
precision of 2.6\% and the efficiencies of triggers
(Section~\ref{s:Trigger}) applied to the signal yield are known with a
precision of 3\%. The efficiencies for the identification and isolation
of leptons are observed to be consistent within 5\% for data and
simulation; we take this difference as an uncertainty. The b-tagging
efficiency has been varied within its uncertainties for b, c, and light
flavor jets, leading to final yield uncertainties within 3\% for all
signal mass points. The systematic uncertainty in signal yield that is
associated with the JES~\cite{jes} is obtained by varying the jet energy
scale within its uncertainty; the final uncertainties for all signal
mass points are within 10\%. Systematic uncertainties in the signal
efficiency due to PDFs have been
calculated~\cite{pdf,pdf_2,Ball:2014uwa}, and are constant at $\sim$5\%. The effect of the systematic uncertainty due to the modeling of ISR
jets by the simulation is studied by deriving data/simulation scale
factors that depend on \njets. The maximum size of these uncertainties
varies between 8 and 10\% for different decay modes.

\subsection{Summary of the single-lepton search}
\label{s:1lepRes}

We develop a \dm-dependent signal selection tool with BDTs for the tt and
bbWW decay modes. For each BDT selection shown in Fig.~\ref{fig:BestBDT}
we provide in Table~\ref{tab:ResBdt} the predicted background yield
(without signal contamination) as well as the number of observed data
events for the BDT selections. We do not observe any excess of data events
compared to the predicted total background.
The background composition varies as function of the different SRs of
various decay modes.
For the tt decay mode, the dominant background is \ttll\ (50-60\% of the
total background) across all SRs.
For the bbWW x=0.25 decay mode, the dominant background is \ttl\ for BDT3,
BDT4, BDT6 (40-55\%), and \ttll\ for BDT1 (58\%). 
For the bbWW x=0.5 decay mode, the dominant background for BDT1 and BDT6 is
\ttll\ (40-70\%), while rare processes dominate for BDT4 and BDT5 ($\sim$80\%).
For BDT3, \ttll\ and rare processes dominate with an equal proportion ($\sim$ 33\%).
For the bbWW x=0.75 decay mode, \ttll\ is the dominant background (45--65\%)
for BDT1 to BDT3, while rare processes dominate for BDT5 (47--61\%).
In Fig.~\ref{fig:BDToutDD075} we show the distribution of the BDT output
for data and the predicted background (without signal contamination) for
two trainings of the bbWW $x=0.75$ case.

The signal contamination is taken into account by calculating a new
estimation of the background in case of signal contamination (see Eqs.
\eqref{eq:sf1} and \eqref{eq:sf2}); this is done separately at each
signal mass point in the \mmplane plane, and for each of the signal
regions defined in Fig.~\ref{fig:BestBDT}. For the calculation of limits
(see Section~\ref{s:FRes}), the number of observed events in data and
expected signal remain the same, while the expected background is
modified to correct for signal contamination in the control regions.
While the effect of this contamination is observed to be almost
negligible at high \dm, it can modify the background estimate up to 25\%
at low \dm.

\begin{table}[!ht]
\centering
\topcaption{Background prediction without signal contamination and observed
data for the BDT selections. The total systematic uncertainties are
reported for the predicted background.}
\resizebox{\textwidth}{!}{
\begin{tabular}{l|ccccccc}
\hline
\text{tt}   &
\text{BDT 1} 	&
\text{BDT 1} 	&
\text{BDT 1} 	&
\text{BDT 2} 	&
\text{BDT 5} 	&
\text{BDT 5} 	& 	\\
&
\text{Low $m(\lsp)$} 	&
\text{Medium $m(\lsp)$} 	&
\text{High $m(\lsp)$} 	&
 	&
\text{Low \dm} 	&
\text{High \dm} 	& 	\\
\hline
\text{Background} 	 & 363 $\pm$ 35 & 46 $\pm$ 16  & 19 $\pm$ 7 & 37 $\pm$ 13 & 6 $\pm$ 2 & 4 $\pm$ 2 & \\
\text{Data} 	         & 286	            & 33	       & 17	        & 33	          & 3	          & 1	          & \\
\hline
\hline
\text{{bbWW ($x=0.25$)}} & \text{BDT 1} 	& \text{BDT 3} 	 & \text{BDT 4} 	        & \text{BDT 4} 	& \text{BDT 6} &  &  \\
                         &	                &                	 & \text{Low $m(\lsp)$}         & \text{High $m(\lsp)$}	&                &  &  \\
\hline
\text{Background} 	 & 42 $\pm$ 11 	& 29 $\pm$ 7 	 & 20 $\pm$ 5 	        & 5 $\pm$ 2  & 6 $\pm$ 3  & & \\
\text{Data} 	         & 27	                & 23	                 & 19	                        & 5      	 & 6	          & &  \\
\hline
\hline
\text{{bbWW ($x=0.50$)}} &
\text{BDT 1} 	&
\text{BDT 1} 	&
\text{BDT 1} 	&
\text{BDT 3} 	&
\text{BDT 4} 	&
\text{BDT 5} 	&
\text{BDT 6} 	\\
&
\text{Low \dm } 	&
\text{Low \dm } 	&
\text{High \dm} 	&
\text{} 	&
\text{} 	&
\text{} 	&
\text{} 	\\
&
\text{Low $m(\lsp)$} 	&
\text{High $m(\lsp)$} 	&
\text{} 	&
\text{} 	&
\text{} 	&
\text{} 	&
\text{} 	\\
\hline
\text{Background} 	 & 14 $\pm$ 5 & 3 $\pm$ 2 & 91 $\pm$ 25 & 7 $\pm$ 2 & 0.8 $\pm$ 0.3 & 0.7 $\pm$ 0.4 & 3 $\pm$ 1  \\
\text{Data} 	         & 16	          & 1	          & 85	            & 4	            & 1	            & 2	            & 5 	 \\
\hline
\hline
\text{{bbWW ($x=0.75$)}} & \text{BDT 1} 	& \text{BDT 2} 	& \text{BDT 3} 	& \text{BDT 5} 	& \text{BDT 5}    & &	\\
                         &               	&               	&               	& \text{Low \dm}      & \text{High \dm} & &	\\
\hline
\text{Background} 	 & 13 $\pm$ 4 	& 23 $\pm$ 7 	 & 11 $\pm$ 3 	& 2 $\pm$ 1 	 & 0.4 $\pm$ 0.2 & &	 \\
\text{Data} 	         & 9	                & 15	                 & 6	                & 3	                 & 0	         & &	 \\
\hline
\end{tabular}
\label{tab:ResBdt}}
\end{table}

\section{Dilepton search}
\label{s:2lep}

\subsection{Selection}
\label{s:2lepSel}

For the three dilepton final states considered in this search ($\Pe\mu$,
$\Pe\Pe$, and $\mu\mu$), we define the preselection as follows:
\begin{itemize}
\item At least two oppositely charged leptons.
\item For the leading and sub-leading lepton, we require $\pt > 20$ and $\pt > 10$\GeV, respectively.
\item For all lepton flavors: $\mll >20$\GeV.
\item If more than two lepton pairs are found that satisfy the above three requirements,
the pair with the highest \pt is chosen.
\item For $\Pe\Pe, \mu\mu$ channels: $\abs{\mz - \mll}>25$\GeV (Z boson veto) and $\met > 40$\GeV.
\item $\njets \geq 2$ and $N(\bjets)\geq 1$.
\end{itemize}
At the preselection level, \ttbar production with two leptons
represents $\sim$90\% of the total expected background.

In this search we separate the signal from the dileptonic \ttbar background by constructing a
transverse mass variable \MTTwoS{\ell}{\ell} as defined in Eq.~\eqref{eq:MT2def}.
We begin with the two selected leptons $\ell_1$ and $\ell_2$.
Under the assumption that the \ptvecmiss originates only from two neutrinos, we partition
the \ptvecmiss into two hypothetical neutrinos with transverse momenta \SpecvecptMiss{1}
and \SpecvecptMiss{2}. We calculate the transverse mass \mt of the pairings of these
hypothetical neutrinos with their respective lepton candidates and record the maximum
of these two \mt. This process is repeated with other viable partitions of the \ptvecmiss
until the minimum of these maximal \mt values is reached; this minimum is the
\MTTwoS{\ell}{\ell} for the event~\cite{mt2,Burns:2008va}:
\begin{equation}
\label{eq:MT2def}
\MTTwoS{\ell}{\ell}= \min_{\SpecvecptMiss{1} + \SpecvecptMiss{2} = \ptvecmiss} \left( \max \left[ \mt(\vecpt^{\,\ell_1},\SpecvecptMiss{1}), \mt(\vecpt^{\,\ell_2},\SpecvecptMiss{2}) \right] \right).
\end{equation}

\begin{figure}
\centering
\includegraphics[width=7.5cm]{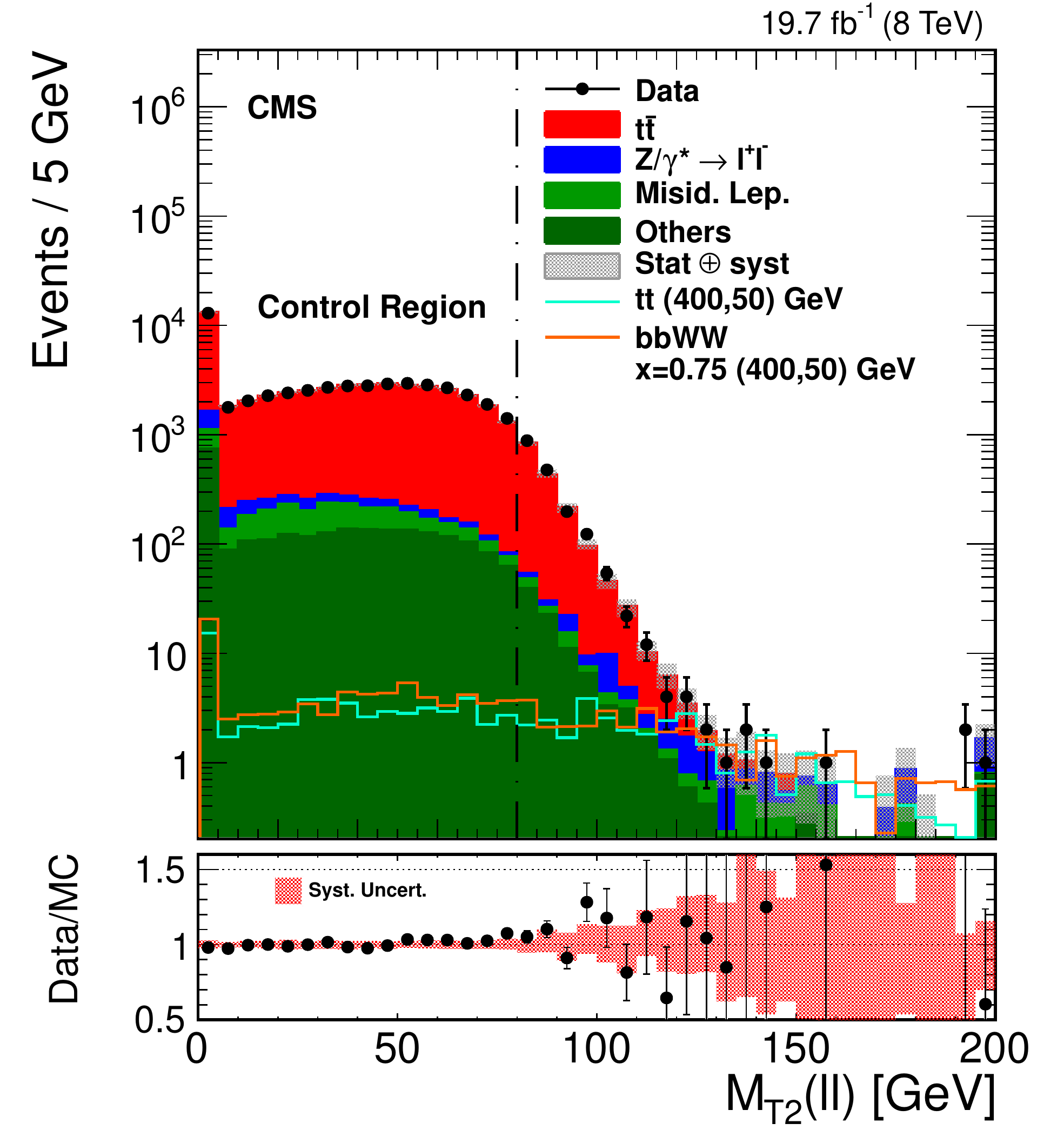}
\includegraphics[width=7.5cm]{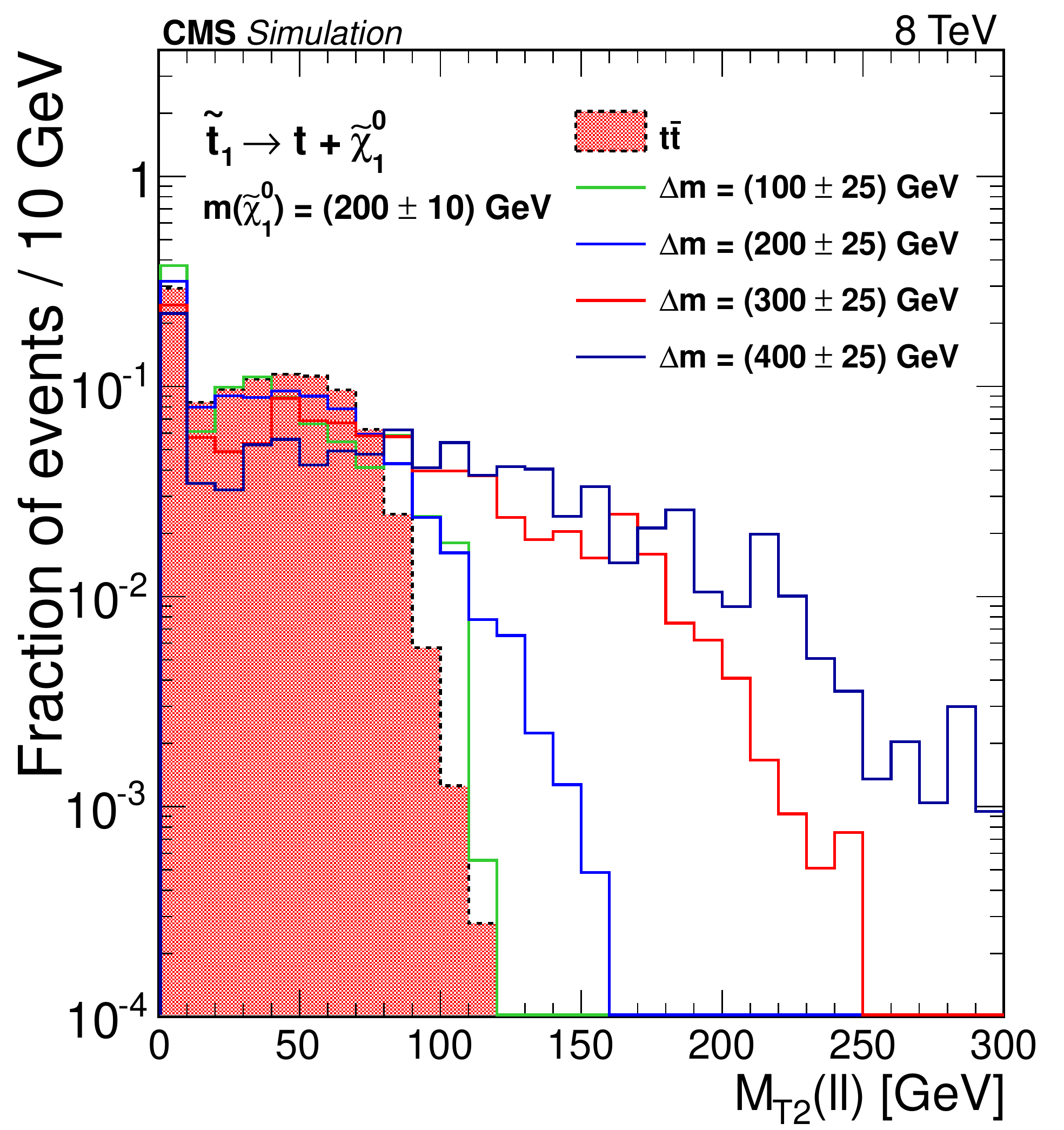}
\caption{Left: Data, expected background, and signal contributions in the
\MTTwoS{\ell}{\ell} distribution at the preselection
level. Background processes are estimated as in Section~\ref{s:2lepBckg}. The uncertainty
bands are calculated from the full list of uncertainties discussed in Section~\ref{s:2lepSys}.
The same signal mass point $\mmplane = (400,50)$\GeV is represented for the
tt and bbWW ($x=0.75$) decay modes.
Right: \MTTwoS{\ell}{\ell} distribution for the \ttbar background and different signal
mass points of the tt decay mode regrouped in constant $\Delta m$ bands;
distributions are normalized to the same area.
}
\label{fig:MET_MT2llPreSelection}
\end{figure}

When constructed in this fashion, \MTTwoS{\ell}{\ell} has the property that
its distribution in \ttll\ events has a kinematic endpoint at $m(\PW)$.
The presence of additional invisible particles for the signal breaks the
assumption that the \ptvecmiss arises from only two neutrinos; consequently, \MTTwoS{\ell}{\ell}
in dileptonic top squark events does not necessarily have an endpoint at $m(\PW)$.
The value of $m(\PW)$ therefore dictates the primary demarcation between the control region
$\MTTwoS{\ell}{\ell} < 80$\GeV, and the general signal region $\MTTwoS{\ell}{\ell} > 80$\GeV.
The left plot of Fig.~\ref{fig:MET_MT2llPreSelection} shows the distribution of
\MTTwoS{\ell}{\ell} at the preselection level, where we observe its discriminating power
for two representative signal mass points.
The distribution of \MTTwoS{\ell}{\ell} in top squark events, however, depends
upon the signal mass point \mmplane, as can be observed on the right plot of
Fig.~\ref{fig:MET_MT2llPreSelection}.

 The optimal threshold on \MTTwoS{\ell}{\ell} for the final selection is thus dependent on
the supersymmetric particle masses:
using the background predictions from Section~\ref{s:2lepBckg} for the \MTTwoS{\ell}{\ell}
signal region, we iterate in 10\GeV steps through possible \MTTwoS{\ell}{\ell}
thresholds, from 80\GeV to 120\GeV; for each \mmplane signal mass point,
we pick the threshold that yields the lowest
expected upper limit for the top squark production cross section,
$\sigma_{95}^\text{exp}$.

\subsection{Background prediction}
\label{s:2lepBckg}

For the \MTTwoS{\ell}{\ell} signal regions used in this search, the dominant
background is \ttbar. Other backgrounds also contribute,
including DY, single-lepton events with
an additional misidentified lepton (see Section~\ref{s:FakeEst}), and rare processes.
The rare processes include single top quarks
produced in association with a W boson; diboson production, including W or Z production
with an associated photon; triple vector boson production; and \ttbar production
in association with one or two vector bosons.
The normalization of the \ttbar and DY backgrounds, and the normalization and shape of the misidentified
lepton backgrounds, are evaluated from data using control samples.
The shapes of the \ttbar and DY backgrounds, and the normalization and
shapes of less common processes, are all estimated from the simulation.
We perform a number of checks to validate the modeling of the \MTTwoS{\ell}{\ell}
distribution in our simulation (see Section~\ref{s:MT2Modeling}).
For the background processes estimated from simulation, we apply the corrective
scale factors mentioned in Sec.~\ref{s:Reco}.

\subsubsection{\texorpdfstring{\ttbar}{ttbar} estimation}
\label{s:ttbar}

The \ttll\ background represents about 90\% of the events in the control region
$\MTTwoS{\ell}{\ell} < 80$\GeV (see Fig.~\ref{fig:MET_MT2llPreSelection} left).
We can therefore use this region to determine the normalization of the expected
SM \ttbar contribution in the signal region. To accomplish this, we first count
the number of data events in the control region and subtract the simulation background
contributions of all non-\ttbar backgrounds; we then normalize it by the simulated \ttbar
yield in the control region. This procedure yields a scale factor of $1.024\pm0.005$.
In this control region, the signal contamination relative to the expected \ttbar contribution
depends upon the \dm considered: While being completely negligible
at high \dm, it can take values between 5\% and 40\% at low \dm, depending on \dm
as well as the considered
top squark decay mode.

\subsubsection{Estimation of the Drell--Yan background}
\label{s:routin}

To estimate the contribution of DY events in the selected events,
we use the Z-boson mass resonance in the \mll distribution
for opposite charge and same flavor dilepton events.
From comparisons with data, we find that our simulation
accurately models the Z mass line shape within systematic uncertainties.
We can therefore calculate a normalization scale factor
for simulated DY events by comparing the observed number
of events inside the Z-veto region ($N^{\ell^{+}\ell^{-}}_{\text{in}}$)
against the expected number of DY events calculated from the simulation
($N^{\text{DY}}_\text{in}$),
\begin{eqnarray}
{SF}^{\ell^{+}\ell^{-}}_{\text{DY}} = \frac{\left(N^{\ell^{+}\ell^{-}}_{\text{in}}- 0.5 N^{\Pe\mu}_{\text{in}} k_{\ell\ell}\right)}{N^{\text{DY}}_\text{in}},
\label{eq:dyest}
\end{eqnarray}
where the number of events with different  flavor ($N^{\Pe\mu}_{\text{in}}$) is subtracted
to account for non-DY processes contaminating $N^{\ell^{+}\ell^{-}}_{\text{in}}$.
The $k$-factors in Eq.~\eqref{eq:dyest} account for different
reconstruction efficiencies for electrons and muons.
Using Eq.~\eqref{eq:dyest}, we calculate a scale factor of ($1.43\pm0.04$)
for $\mu\mu$ events and ($1.46\pm0.04$) for ee events.
To account for the contribution  of $\Pe\mu$ events originating
from $\PZ\to\tau^+\tau^-$ decays, we estimate a scale factor of
($1.44\pm0.04$) for $\Pe\mu$ events by taking the geometric average of the scale factors
for the same-flavor channels.

\subsubsection{Misidentified lepton background estimation}
\label{s:FakeEst}

The misidentified lepton background consists of events in which non-prompt leptons pass
the identification criteria.
The largest category of events falling in this group are semileptonic \ttbar
events and leptonically decaying W events where a jet, or a lepton
within a jet, is misreconstructed as an isolated prompt lepton.

 In order to have an estimation of this background from data, we first measure the
lepton misidentification rate, which is the probability for a non-prompt
lepton to pass the requirements of an isolated lepton. This is done by counting the rate at
which leptons with relaxed identification (``loose'' leptons) pass the ``tight'' selection
requirements (see Section~\ref{s:Reco}). The measurement is performed in a data
sample dominated by multijet events.

We then measure the prompt lepton rate, which is the efficiency for isolated and
prompt leptons to pass selection requirements,
in a data sample enriched in $\Z\to\ell^+\ell^-$ events. As with the misidentification rate,
the prompt rate is determined by counting the rate at which loose leptons pass tight selection
requirements.

Both the measurements of the lepton misidentification rate and the
prompt lepton rate are performed as functions of lepton \pt and
$\abs{\eta}$.
For each dilepton event where both selected leptons pass at least the
loose selection requirements, the measured misidentification and prompt
rates directly translate into a weight for the event. These weights
depend upon whether neither, one, or both loose leptons also passed the
tight selection requirements. The shape and normalization of the
misidentified lepton background is then extracted by first applying
these derived weights to the data sample where both selected leptons
pass at least the loose selection requirements, and then calculating the
weighted distribution of relevant variables such as \MTTwoS{\ell}{\ell}.
Once the background is determined, the number of events falling into the
\MTTwoS{\ell}{\ell} signal regions is found.

\subsection{Checks of the \texorpdfstring{\MTTwoS{\ell}{\ell}}{MT2ll} shape}
\label{s:MT2Modeling}

\begin{figure}
\centering{}
\includegraphics[width=7.5cm]{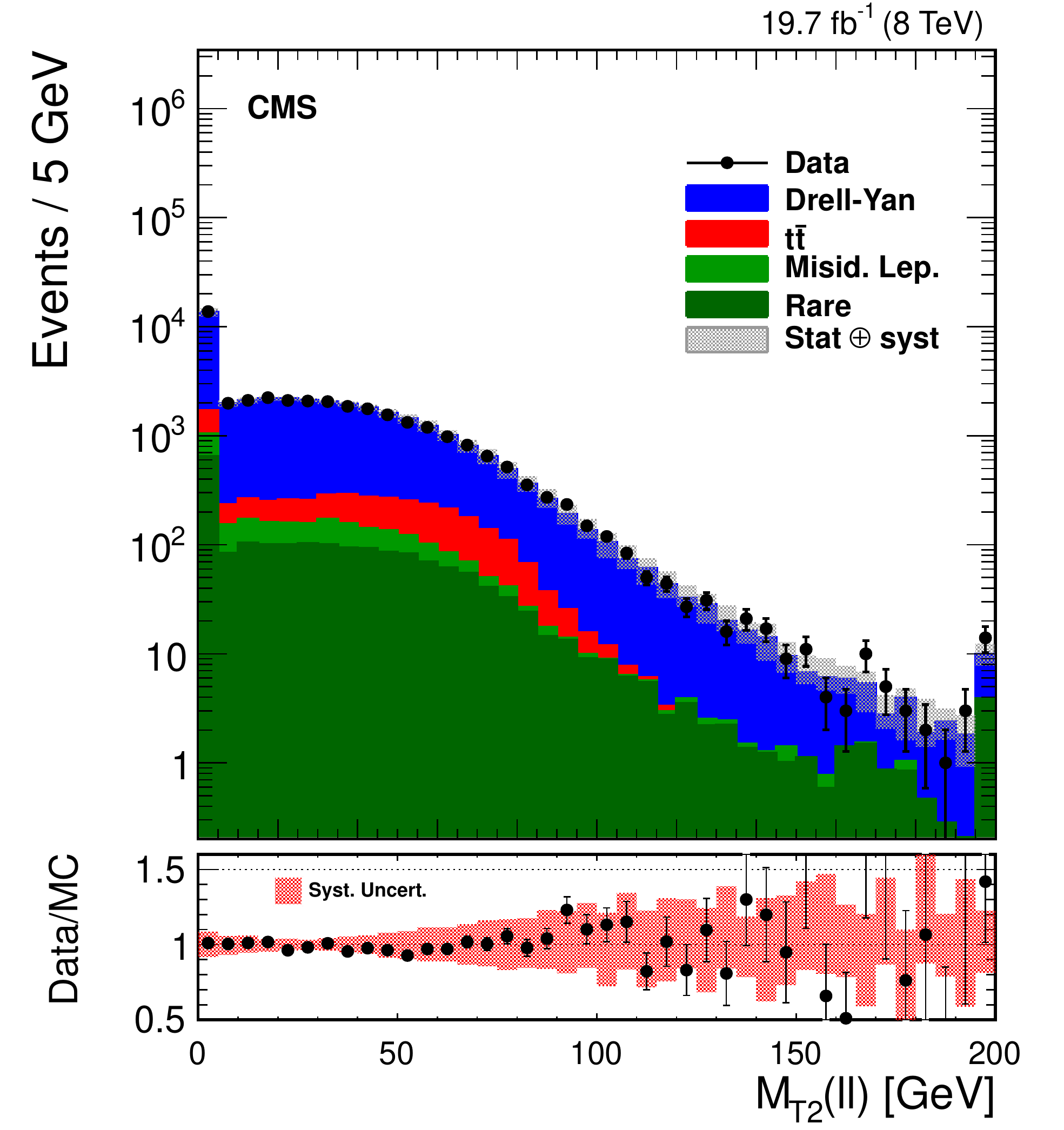}
\caption{Data and expected background contributions for the \MTTwoS{\ell}{\ell} distribution
in a control region enriched in $Z\to\ell\ell$ events. This control region is similar to the preselection,
except that the Z boson veto and \bjet requirements have been inverted.
Background processes are estimated as in Section~\ref{s:2lepBckg}. The uncertainty
bands are calculated from the full list of uncertainties discussed in Section~\ref{s:2lepSys}.
}
\label{fig:MT2ll_ZCR}
\end{figure}

The search in the dileptonic final states requires a good understanding of the
\MTTwoS{\ell}{\ell} shape. In this section we provide a number of validation studies
performed with simulation, with comparisons to data in control regions.

One of the main factors determining the \MTTwoS{\ell}{\ell} shape is the intrinsic
resolution and energy scale of the input objects used in the \MTTwoS{\ell}{\ell} calculation.
From studies using $Z\to\ell\ell$ events, we confirm that the Gaussian core of the \MET
resolution function is sufficiently well-modeled by the simulation. These studies also confirm that
the resolution and scale of the lepton \vecpt are both well-modeled in the simulation.

The intrinsic width of the intermediate W bosons in dileptonic \ttbar
events drives the shape of the \MTTwoS{\ell}{\ell} distribution near the
kinematic edge at 80\GeV. Comparisons of events with different generated
W widths (between 289 MeV and 2.1\GeV) show that any systematic
uncertainty in the W boson width has a negligible effect in the selected
signal regions.

The final notable effect driving the \MTTwoS{\ell}{\ell} shape is the category of events
populating the tails of the \MET resolution function. To confirm that this class of events
is modeled in simulation with reasonable accuracy, we perform comparisons between data and
simulation in a control region enriched in $Z\to\ell\ell$ events; this control region is obtained
by inverting the Z boson veto and requiring zero reconstructed \bjets.
Figure~\ref{fig:MT2ll_ZCR} shows the \MTTwoS{\ell}{\ell} distribution in this control region,
illustrating that the data distribution, including expected events in the tail, is well-modeled
by the simulation.

\subsection{Systematic uncertainties}
\label{s:2lepSys}

We present the dominant systematic uncertainties affecting the dilepton search.

\subsubsection{Systematic uncertainties affecting the background and signal}
\label{s:2lepSysBoth}

The \MET measurement, and subsequently the shape of the \MTTwoS{\ell}{\ell}
distribution, is affected by uncertainties in the lepton energy scale, the JES, the jet energy
resolution, and the scale of the unclustered energy (objects with $\pt<10\GeV$) in the event.
We vary the four-vector momenta of the lepton and jets within their systematic
uncertainties, and propagate the shifted \vecpt back into the \MET and
\MTTwoS{\ell}{\ell} calculations.
For the jet energy resolution uncertainty, we vary it within its uncertainty
and propagate it back into the \MET calculation.
For the unclustered energy scale, we scale the total \vecpt of the unclustered
energy by $\pm10\%$ and propagate it back into the \MET calculation.

As with the single-lepton search (see Section~\ref{s:1lepSys}), we also apply systematic
uncertainties to account for the intrinsic statistical uncertainty in the
simulation samples as well as any mismodeling by the simulation of the b-tagging
efficiency, the lepton trigger efficiency, the lepton ID and isolation, and the limited modeling of
ISR jets by the simulation.
No substantial correlation has been observed between the value of \MTTwoS{\ell}{\ell}
and the size of these four systematic uncertainties.

\subsubsection{Systematic uncertainties affecting only the background}
\label{s:2lepSysBckg}

For the two background normalizations (\ttbar and DY),
we account for the statistical uncertainty in the normalization.
For the misidentified lepton background (see Section~\ref{s:FakeEst}), the two primary sources
of systematic uncertainty are the statistical uncertainty in the measured rates of prompt
and misidentified leptons, and any systematic uncertainty in the measurement of the
misidentification rate. Combining these in quadrature yields a total systematic uncertainty of
$\sim$75\% for the considered signal regions.
For the diboson background processes, which are estimated from the simulation, we
apply a conservative cross section uncertainty of 50\%.

Table~\ref{tab:syst_dilep} displays the magnitude of the effect of the aforementioned
systematic uncertainties (Sections~\ref{s:2lepSysBoth} and \ref{s:2lepSysBckg}) on the
background estimate for each of the considered signal regions.

\begin{table}
\centering
\topcaption{The relevant sources of systematic uncertainty in the
background estimate for each signal region used in the limit setting.
From left to right, the systematic uncertainty sources are: lepton
energy scale ($\ell$ ES), jet energy scale (JES), unclustered energy scale (Uncl.), \MET energy
resolution from jets (JER), uncertainty in b tagging scale factors (b tag), lepton
selection efficiency ($\ell$ eff.), ISR reweighting (ISR), the misidentified lepton estimate
(ML), and the combined normalization uncertainty in the \ttbar, DY, and other
electroweak backgrounds ($\sigma$).}
\resizebox{\textwidth}{!}{
 \begin{tabular}{ l | l  l  l  l  l  l  l  l  l  l  l }
 \hline
\multicolumn{1}{c}{\multirow{2}{*}{\MTTwoS{\ell}{\ell}}}&  \multicolumn{11}{c@{\protect\hphantom{$\,^{+}_{-}\,$}}}{{Systematic uncertainties (\%)}} \\
\cline{2-12}
\multicolumn{1}{c}{}& \multicolumn{1}{r@{\protect\hphantom{$\,^{+}_{-}\,$}}}{Stat.} & 
\multicolumn{1}{r}{$\ell$ ES} & \multicolumn{1}{r}{JES} & \multicolumn{1}{r}{Uncl.} & \multicolumn{1}{r}{JER} & \multicolumn{1}{r}{b tag} & \multicolumn{1}{r}{$\ell$ eff.} & \multicolumn{1}{r}{ISR} & \multicolumn{1}{r}{ML} & \multicolumn{1}{r}{$\sigma$} & \multicolumn{1}{r}{Total} \\
\hline
{$\geq$80\GeV}
 & \rule{0pt}{3ex}\rule[-1.5ex]{0pt}{1.5ex} ${\pm1}$ & $_{-5}^{+4}$ & $_{-1}^{+2}$ & $_{-1}^{+3}$ & $_{-3}^{+3}$ & $_{-0}^{+1}$ & $_{-1}^{+1}$ & $_{-1}^{+1}$ & $_{-1}^{+1}$ & $_{-1}^{+1}$ & $_{-6}^{+7}$ \\
{$\geq$90\GeV}
 & \rule{0pt}{3ex}\rule[-1.5ex]{0pt}{1.5ex} $\pm$2 & $_{-6}^{+6}$ & $_{-2}^{+5}$ & $_{-1}^{+7}$ & $_{-4}^{+7}$ & $_{-0}^{+2}$ & $_{-1}^{+1}$ & $_{-0}^{+0}$ & $_{-2}^{+2}$ & $_{-1}^{+1}$ & $_{\hphantom{0}-9}^{~+14}$ \\
{$\geq$100\GeV}
 & \rule{0pt}{3ex}\rule[-1.5ex]{0pt}{1.5ex} $\pm$4 & $_{-5}^{+6}$ & $_{-2}^{+9}$ & $_{-1}^{+10}$ & $_{\hphantom{0}-2}^{~+12}$ & $_{-1}^{+1}$ & $_{-1}^{+1}$ & $_{-1}^{+2}$ & $_{-3}^{+3}$ & $_{-2}^{+2}$ & $_{\hphantom{0}-9}^{~+20}$ \\
{$\geq$110\GeV}
 & \rule{0pt}{3ex}\rule[-1.5ex]{0pt}{1.5ex} $\pm$7 & $_{-5}^{+9}$ & $_{-1}^{+9}$ & $_{-0}^{+4}$ & $_{-0}^{+5}$ & $_{-2}^{+1}$ & $_{-0}^{+0}$ & $_{-2}^{+3}$ & $_{-7}^{+7}$ & $_{-5}^{+5}$ & $_{-13}^{+18}$ \\
{$\geq$120\GeV}
 & \rule{0pt}{3ex}\rule[-1.5ex]{0pt}{1.5ex} $\pm$10 & $_{-5}^{+4}$ & $_{\hphantom{0}-3}^{~+12}$ & $_{-0}^{+2}$ & $_{-0}^{+5}$ & $_{-1}^{+3}$ & $_{-0}^{+0}$ & $_{-4}^{+6}$ & $_{-12}^{+12}$ & $_{-5}^{+5}$ & $_{-18}^{+22}$ \\
\hline
\end{tabular}
}
\label{tab:syst_dilep}
\end{table}

\subsubsection{Systematic uncertainties affecting only the signal}

As in the single-lepton search, we account for the effect of PDF uncertainties in
the signal efficiency. The resulting uncertainty in signal efficiency is found to be $\sim$4\%
across all signal mass points.

\subsection{Summary of the dilepton search}
\label{s:2lepRes}

We have developed a signal selection based on the \MTTwoS{\ell}{\ell} distribution.
Table~\ref{tab:BckgDat2lep} presents the predicted backgrounds as
well as the number of observed data events for all signal regions; we do not
observe any excess of data events compared to the predicted total background.
Top quark pair production dominates the composition of the total predicted background in the four
signal regions with the lowest \MTTwoS{\ell}{\ell} threshold, decreasing from
91 \% to 45\% with increasing threshold, while DY dominates in the last region
($\sim$38\%).
As with the single-lepton search, the signal contamination is also taken into
account in the final interpretation of the results.

\begin{table}[!ht]
\centering
\topcaption{
Data yields and background expectation for five different \MTTwoS{\ell}{\ell} threshold values.
The asymmetric uncertainties quoted for the background indicate the total systematic
uncertainty, including the statistical uncertainty in the background expectation.}
{
\begin{tabular}{l|ccccc}
\hline
\MTTwoS{\ell}{\ell} threshold & $80\GeV$  & $90\GeV$ & $100\GeV$ & $110\GeV$ & $120\GeV$ \\
\hline
Data & 1785 & 427 & 106 & 30 & 14 \\
 Expected background & 1670 $\,^{+117}_{-104}\,$ & 410 $\,^{+55}_{-35}\,$ & 100 $\,^{+20}_{-8}\,$ & 31.8 $\,^{+5.8}_{-4.0}\,$ & 14.8 $\,^{+3.3}_{-2.7}\,$ \\
\hline
\end{tabular}
}
\label{tab:BckgDat2lep}
\end{table}

\section{Combination and final results}
\label{s:FRes}

After applying all selections for the single-lepton and dilepton data
sets, no evidence for direct top squark production is observed (see
Tables~\ref{tab:ResBdt} and \ref{tab:BckgDat2lep}). We proceed to
combine the results of the two searches. In this combination, no overlap
is expected in the event selections of the two searches, and none is
observed. Since the background predictions are primarily based on data
in the two searches, the corresponding systematic uncertainties are
taken to be uncorrelated. Systematic uncertainties affecting the
expected signal, as well as those due to luminosity, b tagging, PDF,
JES, and lepton identification and isolation, are treated as 100\%
correlated between the two searches.

\begin{figure}[!hbt]
\centering
\includegraphics[width=0.49\textwidth]{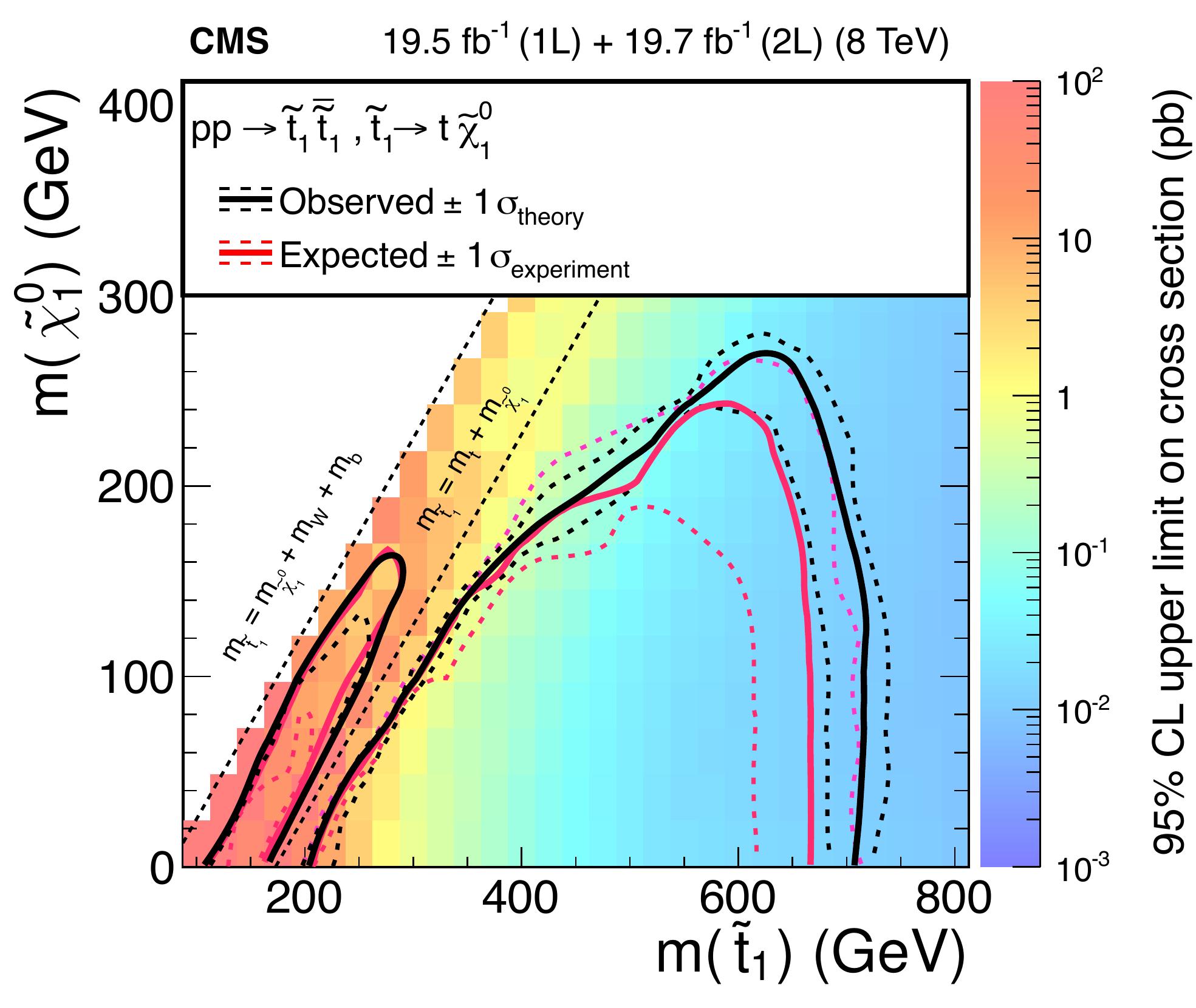}
\includegraphics[width=0.49\textwidth]{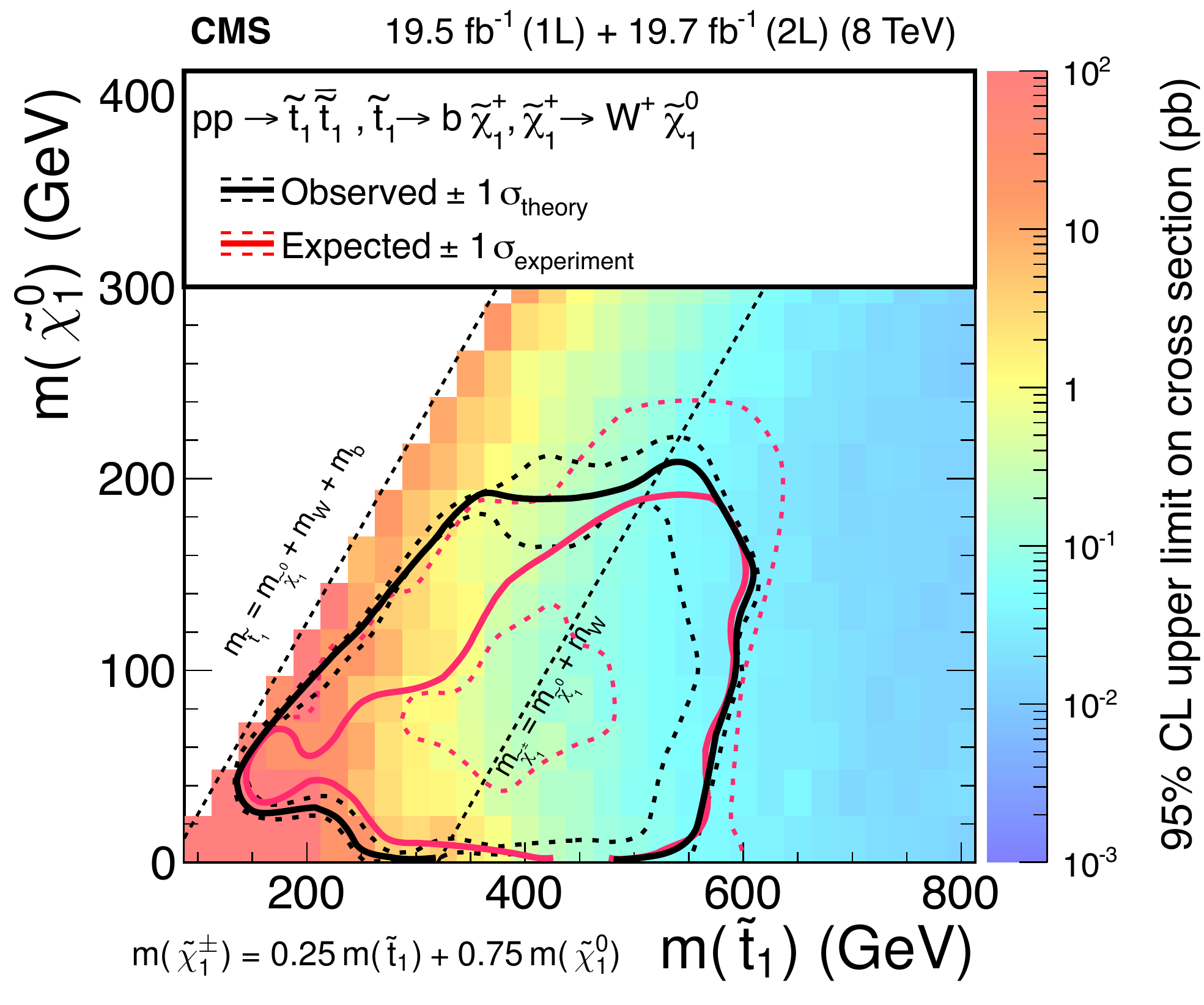} \\
\includegraphics[width=0.49\textwidth]{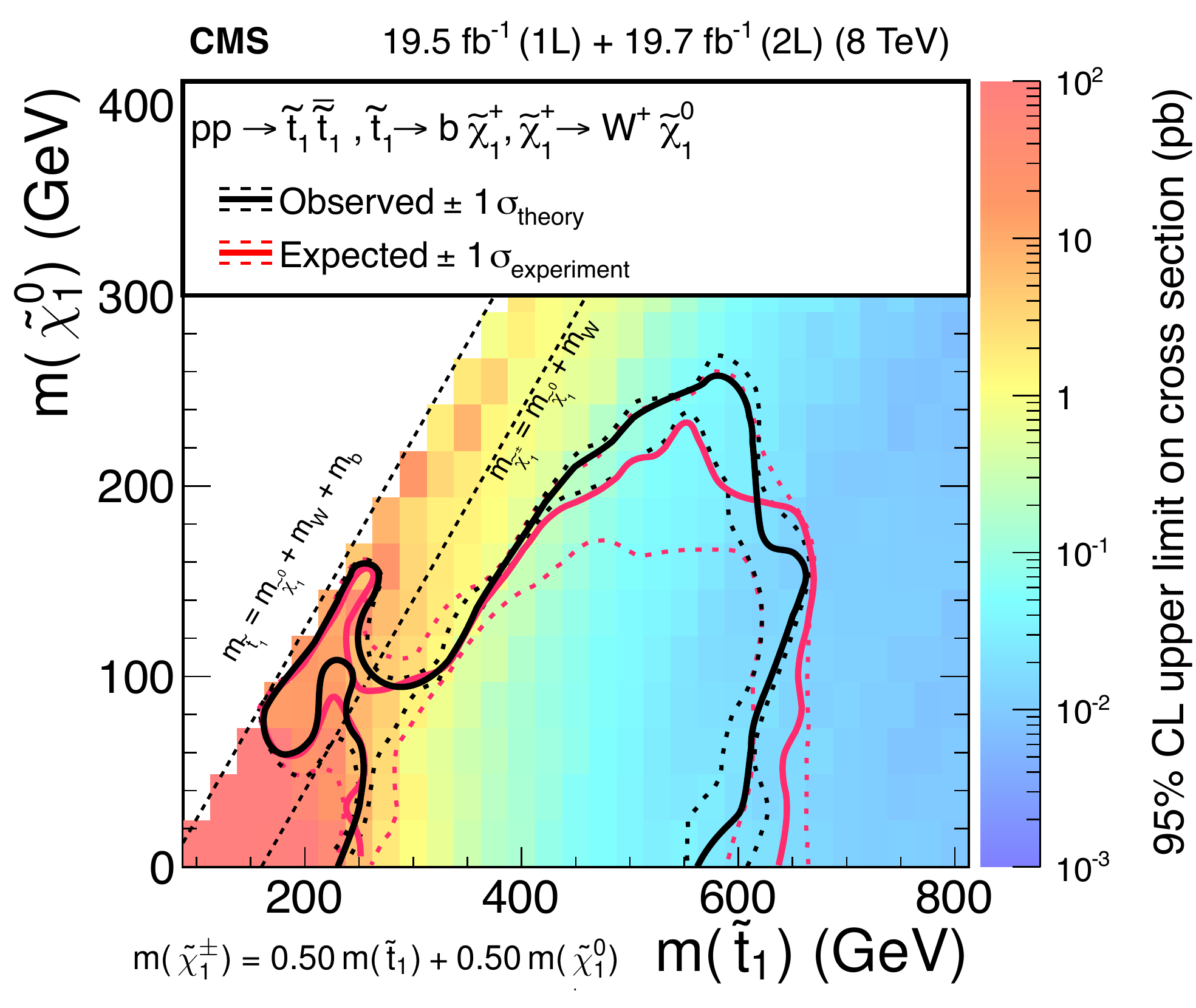}
\includegraphics[width=0.49\textwidth]{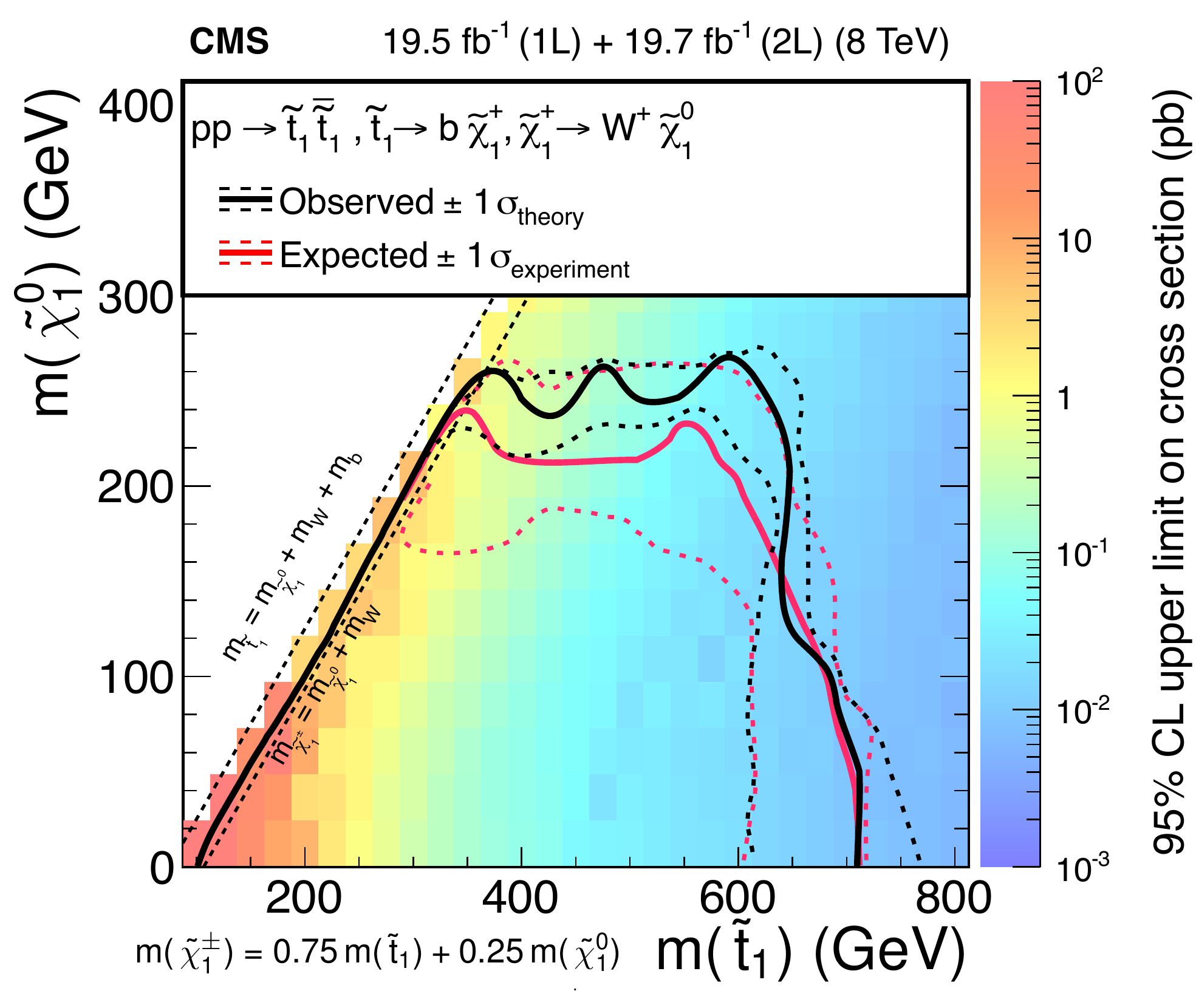} \\
\caption{
\label{fig:bdt_limit_sc}
Exclusion limit at 95\% CL obtained with a statistical combination of the
results from the single-lepton and dilepton searches, for the tt (top left),
bbWW $x=0.25$ (top right), bbWW $x=0.50$ (bottom left) and bbWW $x=0.75$ (bottom
right) decay modes. The red and black lines represent the expected and observed
limits, respectively; the dotted lines represent in each case the $\pm$1 $\sigma$
variations of the contours. For all decay modes, we show the kinematic limit
$m(\stp)=m(\cPqb)+m(\PW)+m(\lsp)$ on the left side of the $(m(\stp),m(\lsp))$ plane; for
the tt decay mode, we show the $\dm=m(\cPqt)$ line; and for the bbWW decay mode, we
show the $m(\PSGcpmDo)-m(\lsp)=m(\PW)$ line.}
\end{figure}

We interpret the absence of excess in both single-lepton and dilepton
searches in terms of a 95\% confidence level (CL) exclusion of top
squark pair production in the \mmplane plane. A frequentist CL$_{\rm
S}$ method~\cite{cls1,cls2,cls3} with a one-sided profile is used,
taking into account the predicted background and observed number of data
events, and the expected signal yield for all signal points.
In this method, Poisson likelihoods are assigned to each of the single-lepton
and dilepton yields, for each \mmplane signal point, and multiplied to give
the combined likelihood for both observations. The final yields of each analysis
are taken from the signal region corresponding to the considered signal point.
Systematic uncertainties are included as nuisance parameter distributions.
A test statistic defined to be the likelihood ratio between the background only and
signal plus background hypotheses is used to set exclusion
limits on top squark pair production; the distributions of these test
statistics are constructed using simulated experiments.
When interpreting the results for the tt and bbWW decay modes, we make the
hypothesis of unit branching fractions, ${\cal B}(\kern0.2em\sTop_1
\to \cPqt^{(*)} \PSGczDo)=1$ and ${\cal
B}(\kern0.2em\sTop_1 \to \cPqb \PSGcpmDo)=1$,
respectively. The expected and observed limits, for which we combine the
results of both searches and account for signal contamination, are
reported in Fig.~\ref{fig:bdt_limit_sc}; the experimental uncertainties
are reported on the expected contour, while the PDF uncertainty for the
signal cross section, quadratically added to the systematic
uncertainties in 2$\mu_r$ and $\mu_r/2$ renormalization scales of the
top squark pair production cross section, are reported on the observed
contour.

For the tt decay mode, we reach sensitivity up to $m(\stp)\sim700$\GeV
for \lsp mass up to $\sim$250\GeV; there is a loss of sensitivity along
the line $\dm=m(\cPqt)$, which delineates two different scenarios within
the tt decay mode (see Table~\ref{tab:StopDecay}) and where the signal
acceptance drops dramatically. For the bbWW decay mode, the sensitivity
reached in this study ranges from 600 to $\sim$700\GeV in $m(\stp)$,
depending on the values of $m(\lsp)$ and m($\PSGcpmDo$); the
sensitivity is greater in the case of a large
$m(\PSGcpmDo)-m(\lsp)$ mass difference as for $x=0.75$, where
the decay products of the two produced W bosons are more energetic. In
the case of $x=0.50$, there is a drop in sensitivity for
$m(\PSGcpmDo)-m(\lsp)\sim m(\PW)$, which corresponds to the
limit in which the W boson is virtual. Because of the rather low
threshold achievable in lepton \pt, sensitivity extends down to the
kinematic limit $\dm\sim m(\cPqb)+m(\PW)$ for the bbWW $x=0.50$ and 0.75
cases.

The final results are dominated by the single-lepton search, where the
selection is based on a multivariate selection with new discriminating
variables, which is adapted to the kinematics of expected signal events,
and where the discriminating power of selection variables is
quantitatively assessed. The new signal selection presented in this
paper leads to the strengthening and further improvement of the results
of Ref.~\cite{MtPub}. We now account for systematic uncertainties due to
PDFs, and more thoroughly assess the effects of signal contamination.
The combination with the dilepton search extends the sensitivity by
$\sim$25\GeV in the tt decay mode in the $\dm\gtrsim m(\cPqt)$ region,
and in the bbWW ($x=0.50$) decay mode across the
$m(\PSGcpmDo)-m(\lsp)=m(\PW$) region; it very moderately
extends the sensitivity in the bbWW ($x=0.75$) at both high \stp and
\lsp masses; no gain of sensitivity is observed in the bbWW ($x=0.25$)
case where the search is limited by the small
$m(\PSGcpmDo)-m(\lsp)$ mass difference, leaving a rather
limited phase space to the decay products of the $\PW$ boson. The signal
contamination (see Section~\ref{s:1lepRes}) reduces the sensitivity of
the search by 0--30\GeV depending on the decay mode and signal point
under consideration. The limits are rather insensitive to the choice of
hypothesis for the polarization of the interaction in the $\cPqt$\lsp
and $\PW$\lsp\chg couplings for the tt and bbWW decay modes,
respectively.

\section{Conclusions}
\label{s:conc}

Using up to 19.7\fbinv of pp collision data taken at $\sqrt{s}=8$\TeV,
we search for direct top squark pair production in both single-lepton
and dilepton final states. In both searches the standard model
background, dominated by the \ttbar process, is predicted using control
samples in data. In this single-lepton search, we improve the results of
Ref.~\cite{MtPub} by employing an upgraded multivariate tool for signal
selection, fed by both kinematic and topological variables and
specifically trained for different decay modes and kinematic regions.
This systematic approach to the signal selection, where the
discriminating power of each selection variable is quantitatively
assessed, is a key feature of the single-lepton search. The background
determination method has also been improved compared to
Ref.~\cite{MtPub}. In the dilepton search the signal selection is based
on the \MTTwoS{\ell}{\ell}\ variable. In both searches, the effect of
the signal contamination is accounted for. No excess above the predicted
background is observed in either search. Simplified models (Fig. 1) are
used to interpret the results in terms of a region in the \mmplane
plane, excluded at 95\% CL. We combine the results of both searches for
maximal sensitivity; the sensitivity depends on the decay mode, and on
the \mmplane signal point. The highest excluded \stp and \lsp masses
are about 700\GeV and 250\GeV, respectively.

\begin{acknowledgments}
\label{sec:acknowledgements}
\hyphenation{Bundes-ministerium Forschungs-gemeinschaft Forschungs-zentren} We congratulate our colleagues in the CERN accelerator departments for the excellent performance of the LHC and thank the technical and administrative staffs at CERN and at other CMS institutes for their contributions to the success of the CMS effort. In addition, we gratefully acknowledge the computing centres and personnel of the Worldwide LHC Computing Grid for delivering so effectively the computing infrastructure essential to our analyses. Finally, we acknowledge the enduring support for the construction and operation of the LHC and the CMS detector provided by the following funding agencies: the Austrian Federal Ministry of Science, Research and Economy and the Austrian Science Fund; the Belgian Fonds de la Recherche Scientifique, and Fonds voor Wetenschappelijk Onderzoek; the Brazilian Funding Agencies (CNPq, CAPES, FAPERJ, and FAPESP); the Bulgarian Ministry of Education and Science; CERN; the Chinese Academy of Sciences, Ministry of Science and Technology, and National Natural Science Foundation of China; the Colombian Funding Agency (COLCIENCIAS); the Croatian Ministry of Science, Education and Sport, and the Croatian Science Foundation; the Research Promotion Foundation, Cyprus; the Ministry of Education and Research, Estonian Research Council via IUT23-4 and IUT23-6 and European Regional Development Fund, Estonia; the Academy of Finland, Finnish Ministry of Education and Culture, and Helsinki Institute of Physics; the Institut National de Physique Nucl\'eaire et de Physique des Particules~/~CNRS, and Commissariat \`a l'\'Energie Atomique et aux \'Energies Alternatives~/~CEA, France; the Bundesministerium f\"ur Bildung und Forschung, Deutsche Forschungsgemeinschaft, and Helmholtz-Gemeinschaft Deutscher Forschungszentren, Germany; the General Secretariat for Research and Technology, Greece; the National Scientific Research Foundation, and National Innovation Office, Hungary; the Department of Atomic Energy and the Department of Science and Technology, India; the Institute for Studies in Theoretical Physics and Mathematics, Iran; the Science Foundation, Ireland; the Istituto Nazionale di Fisica Nucleare, Italy; the Ministry of Science, ICT and Future Planning, and National Research Foundation (NRF), Republic of Korea; the Lithuanian Academy of Sciences; the Ministry of Education, and University of Malaya (Malaysia); the Mexican Funding Agencies (CINVESTAV, CONACYT, SEP, and UASLP-FAI); the Ministry of Business, Innovation and Employment, New Zealand; the Pakistan Atomic Energy Commission; the Ministry of Science and Higher Education and the National Science Centre, Poland; the Funda\c{c}\~ao para a Ci\^encia e a Tecnologia, Portugal; JINR, Dubna; the Ministry of Education and Science of the Russian Federation, the Federal Agency of Atomic Energy of the Russian Federation, Russian Academy of Sciences, and the Russian Foundation for Basic Research; the Ministry of Education, Science and Technological Development of Serbia; the Secretar\'{\i}a de Estado de Investigaci\'on, Desarrollo e Innovaci\'on and Programa Consolider-Ingenio 2010, Spain; the Swiss Funding Agencies (ETH Board, ETH Zurich, PSI, SNF, UniZH, Canton Zurich, and SER); the Ministry of Science and Technology, Taipei; the Thailand Center of Excellence in Physics, the Institute for the Promotion of Teaching Science and Technology of Thailand, Special Task Force for Activating Research and the National Science and Technology Development Agency of Thailand; the Scientific and Technical Research Council of Turkey, and Turkish Atomic Energy Authority; the National Academy of Sciences of Ukraine, and State Fund for Fundamental Researches, Ukraine; the Science and Technology Facilities Council, UK; the US Department of Energy, and the US National Science Foundation.

Individuals have received support from the Marie-Curie programme and the European Research Council and EPLANET (European Union); the Leventis Foundation; the A. P. Sloan Foundation; the Alexander von Humboldt Foundation; the Belgian Federal Science Policy Office; the Fonds pour la Formation \`a la Recherche dans l'Industrie et dans l'Agriculture (FRIA-Belgium); the Agentschap voor Innovatie door Wetenschap en Technologie (IWT-Belgium); the Agence Nationale de la Recherche ANR-12-JS05-002-01 (France); the Ministry of Education, Youth and Sports (MEYS) of the Czech Republic; the Council of Science and Industrial Research, India; the HOMING PLUS programme of the Foundation for Polish Science, cofinanced from European Union, Regional Development Fund; the OPUS programme of the National Science Center (Poland); the Compagnia di San Paolo (Torino); the Consorzio per la Fisica (Trieste); MIUR project 20108T4XTM (Italy); the Thalis and Aristeia programmes cofinanced by EU-ESF and the Greek NSRF; the National Priorities Research Program by Qatar National Research Fund; the Rachadapisek Sompot Fund for Postdoctoral Fellowship, Chulalongkorn University (Thailand); and the Welch Foundation, contract C-1845.
\end{acknowledgments}
\clearpage

\bibliography{auto_generated}
\cleardoublepage \appendix\section{The CMS Collaboration \label{app:collab}}\begin{sloppypar}\hyphenpenalty=5000\widowpenalty=500\clubpenalty=5000\textbf{Yerevan Physics Institute,  Yerevan,  Armenia}\\*[0pt]
V.~Khachatryan, A.M.~Sirunyan, A.~Tumasyan
\vskip\cmsinstskip
\textbf{Institut f\"{u}r Hochenergiephysik der OeAW,  Wien,  Austria}\\*[0pt]
W.~Adam, E.~Asilar, T.~Bergauer, J.~Brandstetter, E.~Brondolin, M.~Dragicevic, J.~Er\"{o}, M.~Flechl, M.~Friedl, R.~Fr\"{u}hwirth\cmsAuthorMark{1}, V.M.~Ghete, C.~Hartl, N.~H\"{o}rmann, J.~Hrubec, M.~Jeitler\cmsAuthorMark{1}, V.~Kn\"{u}nz, A.~K\"{o}nig, M.~Krammer\cmsAuthorMark{1}, I.~Kr\"{a}tschmer, D.~Liko, T.~Matsushita, I.~Mikulec, D.~Rabady\cmsAuthorMark{2}, N.~Rad, B.~Rahbaran, H.~Rohringer, J.~Schieck\cmsAuthorMark{1}, R.~Sch\"{o}fbeck, J.~Strauss, W.~Treberer-Treberspurg, W.~Waltenberger, C.-E.~Wulz\cmsAuthorMark{1}
% ...
\vskip\cmsinstskip
% ...
1:~~Also at Vienna University of Technology, Vienna, Austria\\
2:~~Also at CERN, European Organization for Nuclear Research, Geneva, Switzerland\\
% ...

\end{sloppypar}
\end{document}